\documentclass[journal,singlecolumn]{IEEEtran}

\usepackage{cite}
\usepackage{amsmath,amssymb,amsfonts,amsthm,array,multirow}
\usepackage{algorithm,algorithmic}
\usepackage[caption=false,font=normalsize,labelfont=sf,textfont=sf]{subfig}
\usepackage{graphicx,tikz,psfrag,pgfplots}
\usepackage[utf8]{inputenc}
\usepackage[english]{babel}
\usepackage{booktabs}
\usepackage{epstopdf}
\usepackage{url}

\usetikzlibrary{plotmarks}
\pgfplotsset{compat=newest}
\pgfplotsset{soldot/.style={color=blue,only marks,mark=*}}
\pgfplotsset{holdot/.style={color=blue,fill=white,only marks,mark=*}}

\newcommand{\rank}{\mathop{\bf rank}}

\newcommand{\prox}{\mathbf{prox}}

\newcommand{\argmin}{\mathop{\rm argmin}}

\newcommand{\sign}{\mathop{\text{sign}}}

\newcommand{\ie}{{\it i.e.}}

\newcommand{\R}{\mathbb{R}}

\newcommand{\vct}[1]{\boldsymbol{#1}}
\newcommand{\mtx}[1]{\boldsymbol{#1}}

\newcommand{\linop}[1]{\mathcal{#1}}	

\newcommand{\vp}{\vct{p}}
\newcommand{\vq}{\vct{q}}
\newcommand{\vr}{\vct{r}}

\newcommand{\vw}{\vct{w}}
\newcommand{\vx}{\vct{x}}
\newcommand{\vy}{\vct{y}}
\newcommand{\vz}{\vct{z}}

\newcommand{\vmu}{\vct{\mu}}

\newcommand{\vphi}{\vct{\phi}}

\newcommand{\vxi}{\vct{\xi}}

\newcommand{\vzero}{\vct{0}}

\newcommand{\mA}{\mtx{A}}
\newcommand{\mB}{\mtx{B}}

\newcommand{\mQ}{\mtx{Q}}

\newcommand{\mW}{\mtx{W}}

\newcommand{\mLambda}{\mtx{\Lambda}}

\newcommand{\mId}{{\bf I}}

\newcommand{\mzero}{{\bf 0}}

\newcommand{\loF}{\linop{F}}

\renewcommand{\sl}{s_\ell}

\newtheorem{prop}{Proposition}

\newcommand{\norm}[1]{\left\lVert#1\right\rVert}

\newcolumntype{L}[1]{>{\raggedright\let\newline\\\arraybackslash\hspace{0pt}}m{#1}}
\newcolumntype{C}[1]{>{\centering\let\newline\\\arraybackslash\hspace{0pt}}m{#1}}
\newcolumntype{R}[1]{>{\raggedleft\let\newline\\\arraybackslash\hspace{0pt}}m{#1}}

\newtheorem{corollary}{Corollary}[]
\newtheorem{remark}{Remark}[]

\begin{document}

\title{A Convex Formulation for Binary Tomography}

\author{Ajinkya~Kadu and
        Tristan~van~Leeuwen
\thanks{A. Kadu and T. van Leeuwen are with the Mathematical Institute,
Utrecht University, The Netherlands.}}


\maketitle

\begin{abstract}
Binary tomography is concerned with the recovery of binary images from a few of their projections (i.e., sums of the pixel values along various directions). To reconstruct an image from noisy projection data, one can pose it as a constrained least-squares problem. As the constraints are non-convex, many approaches for solving it rely on either relaxing the constraints or heuristics. In this paper we propose a novel convex formulation, based on the Lagrange dual of the constrained least-squares problem. The resulting problem is a generalized LASSO problem which can be solved efficiently. It is a relaxation in the sense that it can only be guaranteed to give a feasible solution; not necessarily the optimal one. In exhaustive experiments on small images ($2\times 2$, $3\times 3$, $4\times 4$) we find, however, that if the problem has a unique solution, our dual approach finds it. In case of multiple solutions, our approach finds the commonalities between the solutions. Further experiments on realistic numerical phantoms and an experiment on X-ray dataset show that our method compares favourably to Total Variation and DART.

The code associated with this paper is available at \textcolor{blue}{\url{https://github.com/ajinkyakadu/BinaryTomo}}.
\end{abstract}


\IEEEpeerreviewmaketitle

\section{Introduction}

\IEEEPARstart{D}{iscrete} tomography is concerned with the recovery of discrete images (i.e., images whose pixels take on a small number of prescribed grey values) from a few of their projections (i.e., sums of the pixel values along various directions). Early work on the subject mostly deals with the mathematical analysis, combinatorics, and geometry. Since the 1970s, the development of algorithms for discrete tomography has become an active area of research as well \cite{herman2008advances}. It has found applications in image processing and computer vision \cite{sanz1988radon,sharif2005discrete}, atomic-resolution electron microscopy \cite{san1995structural,carazo1999discrete}, medicine imaging \cite{servieres2003discrete,carvalho1999binary} and material sciences \cite{midgley2009electron,balasko2005neutron,van2011three,batenburg20093d}.


\subsection{Mathematical formulation}
The discrete tomography problem may be mathematically formulated as follows. We represent an image by a grid of $N = n \times n$ pixels taking values $x_j \in \mathcal{U} = \{u_0, u_1, \ldots, u_{K}\}$. The projections are linear combinations of the pixels along $m$ different (lattice) directions. We denote the linear transformation from image to projection data by
\[
\vy = \mA \vx,
\]
where $x_j$ denotes the value of the image in the $j^\text{th}$ cell, $y_i$ is the (weighted) sum of the image along the $i^{\text{th}}$ ray and $a_{ij}$ is proportional to the length of the  $i^{\text{th}}$ ray in the $j^\text{th}$ cell\footnote{We note that other projection models exist and can be similarly represented by $a_{ij}$.}.

The goal is to find a solution to this system of equations with the constraint that $x_j \in \mathcal{U}$, i.e.,
\[
\text{find} \quad  \vx \in \mathcal{U}^N \quad \text{such that} \quad \mA \vx = \vy .
\]
When the system of equations does not have a unique solution, finding one that only takes values in $\mathcal{U}^N$ has been shown to be an NP-hard problem for more than 3 directions, \ie, $m \geq 3$ \cite{Gardner1999}.

Due to the presence of noise, the system of equations may not have a solution, and the problem is sometimes formulated as a constrained least-squares problem
\begin{equation}
\underset{\vx \in \mathcal{U}^N}{\min} \quad \tfrac{1}{2} \|\mA \vx - \vy \|^2.
\label{eq:CLSI}
\end{equation}
Obviously, the constraints are non-convex and solving \eqref{eq:CLSI} exactly is not trivial. Next, we briefly discuss some existing approaches for solving it.

\subsection{Literature Review}
Methods for solving \eqref{eq:CLSI} can be roughly divided into four classes: algebraic methods, stochastic sampling methods, (convex) relaxation and (heuristic/greedy) combinatorial approaches.

The \emph{algebraic} methods exploit the algebraic structure of the problem and may give some insight into the (non-) uniqueness and the required number of projections \cite{yagle1999algebraic, hajdu2001algebraic}. While theoretically very elegant, these methods are not readily generalised to realistic projection models and noisy data.

The \emph{stochastic sampling methods} typically construct a probability density function on the space of discrete images, allowing one to sample images and use Markov Chain Monte Carlo type methods to find a solution \cite{matej1999binary,chan1999probabilistic,frese1999multiscale}. These methods are very flexible but may require a prohibitive number of samples when applied to large-scale datasets.

\emph{Relaxation} methods are based on some form convex or non-convex relaxation of the constraint. This allows for a natural extension of existing variational formulations and iterative algorithms \cite{censor1999binary, vardi1999reconstruction,capricelli2007convex,schule2005discrete,tuysuzoglu2017fast}. While these approaches can often be implemented efficiently and apply to large-scale problems, getting them to converge to the correct binary solution can be challenging. Another variant of convex relaxation include the linear-programming based method \cite{kuske2017novel}. This method works well on small-scale images and noise-free data. 

The \emph{heuristic} algorithms, finally, combine ideas from combinatorial optimization and iterative methods. Such methods are often efficient and known to perform well in practice \cite{Barcucci2001,batenburg2011dart}.

A more extensive overview of various methods for binary tomography and variants thereof (e.g., with more than two grey levels) are discussed in \cite{herman1999discrete}.

\subsection{Contributions and outline}
We propose a novel, convex, reformulation for discrete tomography with two grey values $\{u_0,u_1\}$ (often referred to as \emph{binary} tomography). Starting from the constrained least-squares problem \eqref{eq:CLSI} we derive a corresponding Lagrange dual problem, which is convex by construction. Solving this problem yields an image with pixel values in $\{u_0,0,u_1\}$. Setting the remaining zero-valued pixels to $u_0$ or $u_1$ generates a \emph{feasible} solution of \eqref{eq:CLSI} but not necessarily an \emph{optimal} one. In this sense, our approach is a relaxation. Exhaustive enumeration of small-scale ($n=2, 3, 4$) images with few directions ($m = 2,3$) show that if the problem has a unique solution, then solving the dual problem yields the correct solution. When there are multiple solutions, the dual approach finds the common elements of the solutions, leaving the remaining pixels undefined (zero-valued). We conjecture that this holds for larger $n$ and $m$ as well. This implies that we can only expect to usefully solve problem instances that allow a unique solution and characterize the non-uniqueness when there are a few solutions.
For practical applications, the most relevant setting is where the equations alone do not permit a unique solution, but the constrained problem does. Otherwise, more measurements or prior information about the object would be needed in order to usefully image it. With well-chosen numerical experiments on synthetic and real data, we show that our new approach is competitive for practical applications in X-ray tomography as well.

The outline of the paper is as follows. We first give an intuitive derivation of the dual problem for invertible $\mA$ before presenting the main results for general $\mA$. We then discuss two methods for solving the resulting convex problem. Then, we offer the numerical results on small-scale binary problems to support our conjecture. Numerical results on numerical phantoms and real data are presented in Section IV. Finally, we conclude the paper in Section V.

\section{Dual Problem}
\label{section:dualProblem}
For the purpose of the derivation, we assume that the problem has pixel values are $\pm 1$.
The least-squares \emph{binary} tomography problem can then be formulated as:
\begin{equation}
	\begin{aligned}
		\underset{\vx , \, \vphi }{\min} \quad &
		\tfrac{1}{2}\|\mA\vx - \vy \|^2 \, , \\
		\mbox{subject to} \quad & \vx = \sign (\vphi),
	\end{aligned}
	\label{eq:DT-P:LS}
\end{equation}
where $\vphi \in \mathbb{R}^N$ is an auxiliary variable, and $\sign (\cdot)$ denotes the elementwise signum function. In our analysis, we consider the signum function such that $\sign(0) = 0$. The Lagrangian for this problem is defined as
\begin{align}
	\mathcal{L}(\vx, \vphi, \boldsymbol{\nu}) \triangleq \tfrac{1}{2}\|\mA \vx - \vy \|^2
	+ \boldsymbol{\nu}^T \left( \vx - \sign (\vphi) \right) .
	\label{eq:DT-P:LS:Lag}
\end{align}
The variable $\boldsymbol{\nu} \in \mathbb{R}^N$ is the Lagrange multiplier associated with the equality constraint $\vx = \sign (\vphi)$. We refer to this variable as the \textit{dual} variable in the remainder of the paper. We define the dual function $g(\boldsymbol{\nu})$ corresponding to the Lagrangian \eqref{eq:DT-P:LS:Lag} as
\begin{equation*}
	g(\boldsymbol{\nu}) \triangleq \underset{\vx,\vphi}{\inf} \quad \mathcal{L}(\vx,
	\vphi, \boldsymbol{\nu})
\end{equation*}
The primal problem \eqref{eq:DT-P:LS} has a dual objective expressed as
\begin{equation*}
\underset{\boldsymbol{\nu}}{\max} \quad g(\boldsymbol{\nu}).
\end{equation*}
As the dual function is \emph{always} concave \cite{rockafellar1970convex}, this provides a way to define a convex formulation for the original problem. We should note two important aspects of duality theory here: \emph{i)} we are not guaranteed in general that maximizing $g(\boldsymbol{\nu})$ yields a solution to the primal problem; \emph{ii)} the reformulation is only computationally useful if we can efficiently evaluate $g$. The conditions under which the dual problem yields a solution to the primal problem are known as Slater's conditions \cite{slater1959lagrange} and are difficult to check in general unless the primal problem is convex. We will later show, by example, that the dual problem \emph{does not always} solve the primal problem. Classifying under which conditions we \emph{can} solve the primary problem via its dual is beyond the scope of this paper.

It turns out we can obtain a closed-form expression for $g$. Before presenting the general form of $g$, we first present a detailed derivation for invertible $\mA$ to provide some insight.

\subsection{Invertible $\mA$}
The Lagrangian is \textit{separable} in terms of $\vx$ and $\vphi$. Hence, we can represent the dual function as the sum of two functions, $g_1(\boldsymbol{\nu})$ and $g_2(\boldsymbol{\nu})$.
\begin{align}
	g(\boldsymbol{\nu}) = \underbrace{\underset{\vx}{\inf} \, \big\lbrace \tfrac{1}{2}\|\vy -
	 \mA \vx \|^2 + \boldsymbol{\nu}^T \vx \big\rbrace}_{g_1(\boldsymbol{\nu})} +
	  \underbrace{\underset{\vphi}{\inf} \, \big\lbrace - \boldsymbol{\nu}^T \sign (\vphi)
	  \big\rbrace}_{g_2(\boldsymbol{\nu})}
	  \label{eq:DualFunction}
\end{align}

First, we consider $g_1(\boldsymbol{\nu})$. Setting the gradient to zero we find the unique minimizer:
\begin{align}
\vx^\star &= \left( \mA^T\! \mA \right)^{-1} \left( \mA^T \vy - \boldsymbol{\nu} \right).
\label{eq:DT-P:LS:x}
\end{align}
Substituting $\vx^\star$ back in the expression and re-arranging some terms we arrive at the following expression for $g_1$:
\begin{align}
	g_1(\boldsymbol{\nu}) &=  -\tfrac{1}{2} \| \mA^T\vy - \boldsymbol{\nu} \|_{\left( \mA^T\! \mA
	 \right)^{-1}}^2 + \tfrac{1}{2} \vy^T \vy,
	\label{eq:DT-P:LS:g1}
\end{align}
where $\| \vz \|_{\mW}^2 = \vz^T \mW \vz $ is a weighted $\ell_2$-norm.

Next, we consider $g_2(\boldsymbol{\nu})$. Note that this function is \textit{separable} in terms of $\nu_i$
\begin{align*}
	g_2(\boldsymbol{\nu}) &= \underset{\vphi}{\inf} \, \bigg\lbrace - \sum_{i=1}^{N} \big( \nu_i \sign (\phi_i) \big) \bigg\rbrace.
\end{align*}
The function $- \nu \sign(\phi)$ achieves its smallest value for $\phi = \nu$ when $\nu \not = 0$. This solution is not unique of course, but that does not matter as we are only interested in the sign of $\phi$. When $\nu = 0$ the function takes on value $0$ regardless of the value of $\phi$. We thus find
\begin{align}
g_2(\boldsymbol{\nu}) = -\|\boldsymbol{\nu}\|_1.
\label{eq:Explicitg2}
\end{align}

Hence, the dual function for the Lagrangian in \eqref{eq:DT-P:LS:Lag} takes the following explicit form:
\begin{align}
	g(\boldsymbol{\nu}) &=
	-\tfrac{1}{2} \| \mA^T \vy - \boldsymbol{\nu} \|_{\left( \mA^T\! \mA \right)^{-1}}^2 - \| \boldsymbol{\nu} \|_1
	+ \tfrac{1}{2} \vy^T \vy
	\label{eq:trueDual}
\end{align}
The maximizer to dual function \eqref{eq:trueDual} is found by solving the following minimization problem:
\begin{equation}
	\begin{aligned}
		\underset{\boldsymbol{\nu} \in \mathbb{R}^N}{\min} \quad & \tfrac{1}{2} \|\boldsymbol{\nu}  -\mA^T \vy \|_{\left( \mA^T\!
		 \mA \right)^{-1}}^2 + \| \boldsymbol{\nu} \|_1.
	\end{aligned}
	\label{eq:dual0}
\end{equation}
This optimization problem is famously known as the \emph{least absolute shrinkage and selection operator} (LASSO) \cite{tibshirani1996regression} in the statistics literature. It tries to find a sparse vector in image space by minimizing the distance to the back projection of the data in a scaled $\ell_2$ norm. The primal solution can be synthesized from the solution of the dual problem via $\vx^\star = \sign(\vphi^\star) = \sign(\boldsymbol{\nu}^\star)$.

It is important to note at this point that the solution of the dual problem only determines those elements of the primal problem, $x_i$, for which $\nu_i \not=0$. The remaining degrees of freedom in $\vx$ need to be determined by alternative means. The resulting solution is a \emph{feasible} solution of the primal problem, but not necessarily the \emph{optimal} one.

To gain some insight into the behaviour of the dual objective, consider a one-dimensional example with $\mA = 1$:
\begin{align}
\underset{x \in \{-1,1\} }{\min} \quad & \tfrac{1}{2} (x - y)^2 .
\label{eq:ex:primal}
\end{align}
The solution to this problem is given by $x^\star = \sign(y)$.
The corresponding dual problem is
\begin{align}
\underset{\nu \in \mathbb{R} }{\min} \quad \tfrac{1}{2} (\nu - y)^2 + |\nu|,
\label{eq:ex:dual}
\end{align}
the solution of which is given by $\nu^* = \max(|y| - 1, 0)\sign(y)$. Hence, for $|y| > 1$, the solution of the dual problem yields the desired solution. For $|y| \leq 1$, however, the dual problem yields $\nu^\star=0$ in which case the primal solution $x^* = \sign(\nu^*)$ is not well-defined. We will see in section \ref{sec:sub:solveDual} that when using certain iterative methods to solve the dual problem, the iterations will naturally approach the solution  $\nu^\star=0$ from the correct side, so that the sign of the approximate solution may still be useful.

\subsection{Main results}
We state the main results below. The proofs for these statements are provided in the Appendix section.
\begin{prop}
\label{thm1}
The dual objective of \eqref{eq:DT-P:LS} for general $\mA\in \mathbb{R}^{m\times N}$ is given by
\begin{align*}
g(\boldsymbol{\nu}) =
\begin{cases}
-\tfrac{1}{2} \| \boldsymbol{\nu} - \mA^T \vy\|_{\left( \mA^T \!\mA \right)^{\dagger}}^2 - \| \boldsymbol{\nu} \|_1 + \tfrac{1}{2} \vy^T \vy & \boldsymbol{\nu} \in \mathcal{R}_{\mA}, \\
-\infty & \text{otherwise}
\end{cases}
\end{align*}
where $\dagger$ denotes the pseudo-inverse and $\mathcal{R}_{\mA}$ is the row-space of $\mA$ (i.e., the range of $\mA^T$). This leads to the following optimization problem
\begin{equation}
	\begin{aligned}
		\underset{\boldsymbol{\nu}  \in \mathcal{R}_{\mA} }{\min} \quad & \tfrac{1}{2} \|\boldsymbol{\nu} - \mA^T \vy\|_{\left( \mA^T\!
		 \mA \right)^{\dagger}}^2 + \| \boldsymbol{\nu} \|_1.
	\end{aligned}
	\label{eq:dual1}
\end{equation}
\end{prop}

\begin{remark}
In case $m \geq N$ and $\mA$ has full rank, $\mA^T\!\mA$ is invertible and the general form \eqref{eq:dual1} simplifies to \eqref{eq:dual0}.
\end{remark}

\begin{corollary}
The minimization problem \eqref{eq:dual1} can be restated as
\begin{equation}
	\begin{aligned}
		\underset{\vmu \in \mathbb{R}^m}{\min} \quad & \tfrac{1}{2} \|\mA\mA^\dagger \left(\boldsymbol{\mu} -  \vy\right)\|^2 + \| \mA^T\vmu \|_1,
	\end{aligned}
	\label{eq:dual2}
\end{equation}
and the primal solution is recovered through $\vx^\star = \sign(\mA^T\vmu^\star)$, where $\vmu^\star$ is the solution to \eqref{eq:dual2}.
\label{Corollary:1st}
\end{corollary}

\begin{remark}
For $m \leq N$ and $\mA$ full rank, we have $\mA\mA^\dagger = \mathbf{I}$ and the formulation \eqref{eq:dual2} simplifies to
\begin{align}
		\underset{\vmu \in \mathbb{R}^m}{\min} \quad & \tfrac{1}{2} \| \vmu -  \vy\|^2 + \| \mA^T\vmu \|_1.
		\label{eq:DT:simpleForm}
\end{align}
This form implicitly handles the constraints on the search space of $\boldsymbol{\nu}$ in Proposition~\ref{thm1}. It allows us to use the functional form for matrix $\mA$ thereby reducing the storage and increasing the computational speed to find an optimal dual variable $\vmu^\star$.
\end{remark}

\begin{prop}
The dual problem for a binary tomography problem with grey levels $u_0 < u_1$ is given by:
\begin{equation}
\begin{aligned}
		\underset{\boldsymbol{\nu}  \in \mathcal{R}_{\mA} }{\min} \quad & \tfrac{1}{2} \|\boldsymbol{\nu} - \mA^T \vy\|_{\left( \mA^T\!
		 \mA \right)^{\dagger}}^2 + p(\boldsymbol{\nu}),
	\end{aligned}
	\label{eq:dual3}
\end{equation}
where $p(\boldsymbol{\nu}) = \sum_{i} |u_0|\max(-\nu_i,0) + |u_1|\max(\nu_i,0)$ is an asymmetric one-norm. The primal solution is obtained using
\begin{align*}
\vx^\star = u_0\mathbf{1} + (u_1 - u_0)H(\boldsymbol{\nu}^\star),
\end{align*}
where $H(\cdot)$ denotes the Heaviside function.
\label{Corollary:2nd}
\end{prop}

We summarize the procedure in algorithm~\ref{alg:dual} for finding the optimal solution via solving the dual problem. In practical applications, the formulation in step \ref{alg:dual:lowRank} is very useful since the projection matrix $\mA$ generally has a low rank, \ie, $\rank(\mA) < \min (m,N)$.

\begin{algorithm}
	\caption{Dual problem for various cases}
	\begin{algorithmic}[1]
		\renewcommand{\algorithmicrequire}{\textbf{Input:}}
		\renewcommand{\algorithmicensure}{\textbf{Output:}}
		\REQUIRE $\mA \in \mathbb{R}^{m \times N}$, $\vy \in \mathbb{R}^{m}$
 		\ENSURE  $\vx^\star \in \{ -1, 1\}^N$
 		\IF {$\rank (\mA) = \min (m,N)$}
 			\IF {$m > N$}
 				\STATE $\boldsymbol{\nu}^\star \triangleq \argmin_{\boldsymbol{\nu}} \: \tfrac{1}{2} \| \boldsymbol{\nu} - \mA^T\vy \|_{\left(\mA^T\mA\right)^{-1}}^2 + \| \boldsymbol{\nu} \|_1$
 				\RETURN $\vx^\star = \sign \left( \boldsymbol{\nu}^\star \right)$
 			\ELSE
 				\STATE $\boldsymbol{\nu}^\star \triangleq \argmin_{\boldsymbol{\nu}} \: \tfrac{1}{2} \| \boldsymbol{\nu} - \vy \|^2 + \| \mA^T \boldsymbol{\nu} \|_1$
 				\RETURN $\vx^\star = \sign \left( \mA^T \boldsymbol{\nu}^\star \right)$
 			\ENDIF
 		\ELSE
 			\STATE $\boldsymbol{\nu}^\star \triangleq \argmin_{\boldsymbol{\nu}} \: \tfrac{1}{2} \| \mA \mA^{\dagger}\left( \boldsymbol{\nu} - \vy \right) \|^2 + \| \mA^T \boldsymbol{\nu} \|_1$ \label{alg:dual:lowRank}
 			\RETURN $\vx^\star = \sign \left( \mA^T \boldsymbol{\nu}^\star \right)$
 		\ENDIF	
 	\end{algorithmic}	
 	\label{alg:dual}
\end{algorithm}	

\begin{remark}
The realistic tomographic data contains Poisson noise. In such case, the binary tomography problem takes the constrained weighted least-squares form \cite{fessler1994penalized,sauer1993local}:
\begin{equation}
\begin{aligned}
		\underset{\vx \, , \, \vphi}{\min} \quad & \tfrac{1}{2} \| \boldsymbol{y} -  \mA \vx \|_{\mLambda}^2, \\
		\mbox{subject to} \quad & \vx = \sign (\vphi),
\end{aligned}
\label{eq:BT:weightedLS}
\end{equation}
where $\mLambda \in \mathbb{R}^{m \times m}$ is a diagonal matrix with elements $\Lambda_{i} > 0$ representing the least-squares weight per projection.  The dual objective of \eqref{eq:BT:weightedLS} is given by
\begin{align*}
	g(\boldsymbol{\nu}) = \begin{cases}
	-\tfrac{1}{2} \| \boldsymbol{\nu} - \mA^T \mLambda \vy \|_{\mB}^2 - \| \boldsymbol{\nu} \|_1 +  \tfrac{1}{2} \vy^T \vy & \boldsymbol{\nu} \in \mathcal{R}_{\mA} , \\
	- \infty & \text{otherwise},
	\end{cases} 
\end{align*}
where $\mB \triangleq \left(\mA^T \mLambda \mA \right)^\dagger$. The optimization problem for this dual objective is
\begin{equation}
	\underset{\boldsymbol{\nu} \in \mathcal{R}_{\mA}}{\min} \quad \tfrac{1}{2} \| \boldsymbol{\nu} - \mA^T \mLambda \vy \|_{ \left(\mA^T\mLambda \mA \right)^\dagger}^2 + \| \boldsymbol{\nu} \|_1.
	\label{eq:BT:weightedLS:dual}
\end{equation}
If $\rank (\mA) = m$ with $m \leq N$, the problem \eqref{eq:BT:weightedLS:dual} reduces to
\begin{equation}
	\underset{\vmu \in \mathbb{R}^m}{\min} \quad \tfrac{1}{2} \| \vmu - \mLambda^{1/2} \vy \|^2 + \| \mA^T \mLambda^{1/2} \vmu \|_1,
	\label{eq:BT:weightedLS:dual:simple}
\end{equation}
and the primal solution is recovered from $\vx^\star = \sign ( \mA^T \mLambda^{1/2} \vmu^\star)$, where $\vmu^\star$ is the solution to \eqref{eq:BT:weightedLS:dual:simple}.
\end{remark}

\subsection{Solving the dual problem}
\label{sec:sub:solveDual}
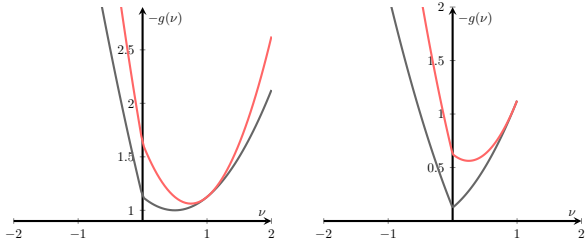
\begin{figure}[t]
	\centering
	\begin{tabular}{cc}
	\begin{tikzpicture}[thick,scale=0.5]
		\begin{axis}[
			samples=100,
			style={ultra thick},
			xlabel={$\nu$},
      		ylabel={$-g(\nu)$},
      		axis y line=middle,
      		axis x line=middle, cycle list name=exotic,
            domain=-2:2, ymin=.9, ymax=2.9, xmin=-2, xmax = 2]
			\addplot[color=black!60] {0.5*(x - 1.5)^2 + abs(x)}; 
			\addplot[color=red!60] {0.5*(1 - 1.5)^2 + 1*(1 - 1.5)*(x - 1) + 0.5*2*(x - 1)^2 + abs(x)}; 
		\end{axis}
	\end{tikzpicture} &
	\begin{tikzpicture}[thick,scale=0.5]
		\begin{axis}[
			samples=100,
			style={ultra thick},
			xlabel={$\nu$},
      		ylabel={$-g(\nu)$},
      		axis y line=middle,
      		axis x line=middle, cycle list name=exotic,
            domain=-1:1, ymin=0, ymax=2, xmin=-2, xmax = 2]
      		\addplot[color=black!60] {0.5*(x - 0.5)^2 + abs(x)}; 
			\addplot[color=red!60] {0.5*(1 - 0.5)^2 + 1*(1 - 0.5)*(x - 1) + 0.5*2*(x - 1)^2 + abs(x)}; 
		\end{axis}
	\end{tikzpicture}
	\end{tabular}
	\caption{Plot of the dual function $g$ (gray line) corresponding the the primal objective $(x - y)^2$ for $y = 1.5$ (left) and $y = 0.5$ (right) and its approximations (red line) at $x = 1$.}
	\label{fig:ex:prox}
\end{figure}

When $\mA^T\!\mA$ is invertible, the dual formulation \eqref{eq:dual1} can be readily solved using a proximal gradient algorithm (\cite{daubechies2004iterative,beck2009fast}):
\begin{equation}
\boldsymbol{\nu}_{k+1} \triangleq S_{L^{-1}}\left( \boldsymbol{\nu}_{k} - L^{-1} \left(\mA^T\!\mA\right)^{-1}\left(\mA^T\vy - \boldsymbol{\nu}_k\right) \right),
\end{equation}
where $L= \|\mA^{-1}\|_2^{2}$ and the soft thresholding operator $S_{\tau}(\cdot) = \max(|\cdot|-\tau,0)\sign(\cdot)$ is applied component-wise to its input. We can interpret this algorithm as minimizing subsequent approximations of the problem, as illustrated in figure \ref{fig:ex:prox}.

An interesting note is that, when starting from $\boldsymbol{\nu}_0 = \mathbf{0}$, the first iteration yields a thresholded version of $\mA^{\dagger}\vy$. As such, the proposed formulation is a natural extension of a naive segmentation approach and allows for segmentation in a data-consistent manner.

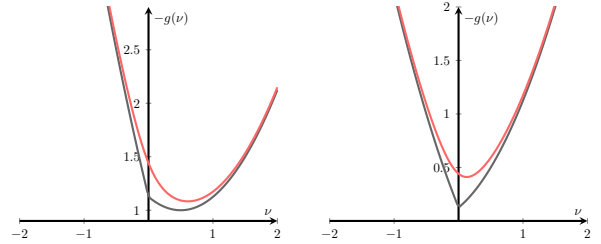
\begin{figure}[t]
	\centering
	\begin{tabular}{cc}
	\begin{tikzpicture}[thick,scale=0.5]
		\begin{axis}[
			samples=100,
			style={ultra thick},
			xlabel={$\nu$},
      		ylabel={$-g(\nu)$},
      		axis y line=middle,
      		axis x line=middle, cycle list name=exotic,
            domain=-2:2, ymin=0.9, ymax=2.9, xmin=-2, xmax = 2]
			\addplot[color=black!60] {0.5*(x - 1.5)^2 + abs(x)}; 
			\addplot[color=red!60] {0.5*(x - 1.5)^2 + sqrt(x^2 + 1e-1)}; 
		\end{axis}
	\end{tikzpicture} &
	\begin{tikzpicture}[thick,scale=0.5]
		\begin{axis}[
			samples=100,
			style={ultra thick},
			xlabel={$\nu$},
      		ylabel={$-g(\nu)$},
      		axis y line=middle,
      		axis x line=middle, cycle list name=exotic,
            domain=-2:2, ymin=0, ymax=2, xmin=-2, xmax = 2]
			\addplot[color=black!60] {0.5*(x - 0.5)^2 + abs(x)}; 
			\addplot[color=red!60] {0.5*(x - 0.5)^2 + sqrt(x^2 + 1e-1)}; 
		\end{axis}
	\end{tikzpicture}
	\end{tabular}
	\caption{Plot of the dual function $g$ (gray line) corresponding the the primal objective $(x - y)^2$ for $y = 1.5$ (left) and $y = 0.5$ (right) and its smooth approximation (red line) using $|\cdot| \approx \sqrt{(\cdot)^2 + \epsilon}$ with $\epsilon = 0.1$.}
	\label{fig:ex:smoothing}
\end{figure}

If $\mA\mA^T$ is invertible we have $\mA\mA^\dagger = \mathbf{I}$ and it seems more natural to solve \eqref{eq:DT:simpleForm} instead. Due to the appearance of $\mA^T$ in the one-norm, it is no longer straightforward to apply a proximal gradient method. A possible strategy is to replace the one-norm with a smooth approximation of it, such as $|\cdot| = \sqrt{(\cdot)^2 + \epsilon}$. As illustrated in figure \ref{fig:ex:smoothing}, this will slightly shift the minimum of the problem. Since we are ultimately only using the sign of the solution, this may not be a problem. The resulting objective is smooth and can be solved using any gradient-descent algorithm. 

We also note that splitting methods can be used to solve \eqref{eq:DT:simpleForm}. For example, the alternating direction method of multipliers (ADMM) \cite{boyd2011distributed} and/or split-Bregman method \cite{goldstein2009split}. Another class of method that can solve \eqref{eq:DT:simpleForm} are the primal-dual methods (e.g., Arrow-Hurwicz primal-dual algorithm \cite{arrow1958studies}, Chambolle-Pock algorithm \cite{chambolle2011first}). These methods rely on the proximal operators of functions and iterate towards finding the saddle point of the problem. If the proximal operators are simple, these are computationally faster than the splitting methods. 

The dual problem \eqref{eq:dual3} for binary tomography problem with grey levels $u_0 < u_1$ is also solved using proximal gradient method. We provide the proximal operator for an asymmetric one-norm in the following proposition.

\begin{prop}
\label{thm:proxForAsyOneNorm}
The proximal operator for an asymmetric one-norm function
\[
	p(\vx) = \sum_{i=1}^{N} |u_0| \max \left( -x_i , 0 \right) + |u_1| \max \left( x_i , 0 \right)
\]
with $u_0 < u_1$, is given by
\[
	\mathcal{P}_{p,\lambda}(\vz) \triangleq \underset{\vx}{\argmin} \left\lbrace \tfrac{1}{2} \| \vx - \vz \|^2 + \lambda p(\vx) \right\rbrace = \mathcal{S}_{\lambda u_0 < \lambda u_1}(\vz),
\]
where $\lambda >0$, and $\mathcal{S}_{a < b}(\cdot)$ is an asymmetric soft-thresholding function
\[
	\mathcal{S}_{a < b}(t) = \begin{cases}
	t - |b| & \quad t \geq |b| \\
	0 & \quad -|a| < t < |b| \\
	t + |a| & \quad t \leq -|a|
	\end{cases}.
\]
\end{prop}

\section{Primal-Dual Algorithm}
We present a proximal-based alternating iterative algorithm to solve~\eqref{eq:DT:simpleForm}. The strong points of the algorithm are (\emph{i}) it does not require matrix inversions, and (\emph{ii}) the convergence parameter can be easily estimated. The method is implemented in the code provided on GitHub (\url{https://github.com/ajinkyakadu/BinaryTomo}).

For simplicity, we rewrite the problem~\eqref{eq:DT:simpleForm} as follows:
\begin{align*}
	\underset{\vmu \in \R^m }{\mbox{minimize}} &\quad h(\vmu) + k(\mA^T \vmu), \\[2ex]
	\mbox{where} \quad & h(\mu) = \tfrac{1}{2} \norm{\vmu - \vy}^2, \qquad k(\vxi) = \norm{\vxi}_1.
\end{align*}
As stated earlier, $h: \R^{m} \mapsto \R$ is a smooth function (\ie, differentiable) while $k: \R^{N} \mapsto \R$ is a non-smooth function. However, the proximal operators for both the functions can be easily computed. To design the algorithm, we look at the first-order optimality condition which reads
\[
	\vzero \in \nabla h(\vmu) + \mA \partial k(\mA^T \vmu) ,
\]
where $\nabla$ denotes the gradient, while $\partial k$ represents the sub-differential for function $k$. Since the first-order optimality condition does not have a closed-form solution, we utilize the splitting technique by introducing $\vz \in \partial k(\mA^T \vmu)$. This results in the following system of equations
\[
	\begin{bmatrix}
	\vzero \\ \vzero 
	\end{bmatrix}
	\in 
	\underbrace{\begin{bmatrix}
	\nabla h & \mA \\
	-\mA^T & \partial k^\star
	\end{bmatrix}	}_{\loF}	
	\underbrace{\begin{bmatrix}
	\vmu \\ \vz
	\end{bmatrix}}_{\vp},
\]  
where $k^\star$ is the (convex) conjugate of the function $k$. Hence, the optimal solution to~\eqref{eq:DT:simpleForm} corresponds to finding a fixed point of $\loF$. It is easy to verify that $\loF$ is a non-expansive monotone operator, \ie,
\[
	\norm{ \loF (\vp) - \loF (\vq) } \leq \norm{\vp - \vq} \qquad \forall \, \vp, \vq.
\]
To find the fixed-point of this non-expansive monotone operator, we use a preconditioned fixed-point method with the preconditioner
\[
	\mQ = \begin{bmatrix}
	\mId /\gamma & - \mA \\
	-\mA^T & \mId/ \gamma
	\end{bmatrix}
\]
where $\gamma \in \left( 0, \|\mA\|^{-1} \right)$ is a parameter that controls the convergence speed of the fixed-point method. The preconditioned fixed-point method produces iterates of form
\begin{align*}
	\left( \loF + \mQ \right) \vp_{t+1} &= \mQ  \vp_{t} \\[1ex]
	\implies \begin{bmatrix}
	\frac{1}{\gamma}\mId + \nabla f & \mzero \\
	-2 \mA^T & \frac{1}{\gamma} \mId + \partial k^\star
	\end{bmatrix}
	\begin{bmatrix}
	\vmu_{t+1} \\ \vz_{t+1}
	\end{bmatrix}
	&=
	\begin{bmatrix}
	\frac{1}{\gamma}\mId & - \mA \\
	-\mA^T & \frac{1}{\gamma}\mId
	\end{bmatrix}
	\begin{bmatrix}
	\vmu_{t} \\ \vz_{t}
	\end{bmatrix}
\end{align*}
\begin{align*}
	\implies 
	\begin{cases}
	\vmu_{t+1} = \left( \mId + \gamma \nabla f \right)^{-1} \left(  \vmu_{t} - \mA \vz_{t} \right) \\
	\vz_{t+1} = \left( \mId + \gamma \partial k^\star \right)^{-1} \left( \vz_{t} - \mA ^T \left( \vmu_{t} - 2 \vmu_{t+1}\right) \right) \\
	\end{cases}
\end{align*}
Noting that $\left( \mId + \gamma \nabla f \right)^{-1}$ denotes a proximal operator of function $f$ while $\left( \mId + \gamma \partial k^\star \right)^{-1}$ is a proximal operator of convex conjugate of function $k$, the primal-dual algorithm results in the following alternating scheme
\begin{align*}
	\mbox{for} \quad t =1, \dots T & \\
	\vmu_{t+1} &= \prox_{\gamma f} \left(  \vmu_{t} - \mA \vz_{t} \right), \\
	\vz_{t+1} &= \prox_{\gamma k^\star} \left( \vz_{t} - \mA ^T \left( \vmu_{t} - 2 \vmu_{t+1}\right) \right),
\end{align*}
where $T$ is the maximum number of iterations. The primal-dual algorithm is simply evaluating the proximal with respect to the primal function $f$ and then the proximal of the dual function $k^\star$ in an alternate fashion. Algorithm~\ref{alg:primal-dual} describes the computational steps in the primal-dual algorithm for solving~\eqref{eq:DT:simpleForm}. Note that the optimal solution can be retrieved using the dual variable $\vz$ as well, as shown in the Algorithm~\ref{alg:primal-dual}.

\begin{algorithm}[t]
	\caption{Primal-Dual algorithm for solving~\eqref{eq:DT:simpleForm}}
	\begin{algorithmic}[1]
		\renewcommand{\algorithmicrequire}{\textbf{Input:}}
		\renewcommand{\algorithmicensure}{\textbf{Output:}}
		\REQUIRE $\mA \in \mathbb{R}^{m \times N}$, $\vy \in \mathbb{R}^{m}$
 		\ENSURE  $\vx^\star \in \{ -1, 1\}^N$
 		\STATE initialize $\vmu_0 = \vzero, \vz_0 = \vzero$
 		\STATE set $\gamma = 0.95/ \norm{\mA} $
 		\FOR {$t=0$ to $T$}
 		\STATE update $\vmu$ \\ $\vmu_{t+1} = \left( \vmu_t - \mA \vz_t + \gamma \vy \right)/(1 + \gamma)$
 		\STATE update $\vz$ \\ 
 			$\hat{\vz}_{t+1} = \vz_{t} - \mA ^T \!  \left( \vmu_{t} - 2 \vmu_{t+1}\right)  $ \\
 			$ \widetilde{\vz}_{t+1} = \max \left(\vzero, \hat{\vz}_{t+1} - 1  \right) - \max \left( \vzero, - \hat{\vz}_{t+1} - 1  \right) $ \\
 			$\vz_{t+1} = \hat{\vz}_{t+1} - \widetilde{\vz}_{t+1}$
 		\STATE check optimality conditions
 		\ENDFOR
 		\STATE $\vx^\star = \sign \left( \vz_T \right)$
 	\end{algorithmic}	
 	\label{alg:primal-dual}
\end{algorithm}	

\subsection{Proximal Operators}
The proximal operator for $h(\vmu) = (1/2)\norm{\vmu - \vy }^2$ is 
\begin{align*}
	\prox_{\gamma h}(\vw) &= \argmin_{\vmu \in \R^{m}} \left\{ \tfrac{1}{2 \gamma} \norm{ \vmu - \vw}^2 +  \tfrac{1}{2} \norm{\vmu - \vy }^2 \right\}, \\
	&= \frac{\vw + \gamma \vy }{1 + \gamma },
\end{align*}
where the $\argmin$ is computed from the setting the gradient of the cost function with resepct to $\vmu$ to zero. To compute the proximal operator of the function $k^\star$, we use the Moreau's identity
\begin{align*}
	\prox_{\gamma k^\star} \! (\vr) = \vr - \gamma \prox_{k / \gamma} (\vr/\gamma).
\end{align*}
Moreau's identity provides a powerful and concise way to compute the proximal of conjugate function from the proximal operator of the function, and vice versa. Although the proximal of the function $k(\vr) = \norm{ \vr}_1$ is trivial, but compute it here for sake of completeness:
\begin{align*}
	\prox_{\gamma h}(\vw) &= \argmin_{\vr \in \R^{N}} \left\{ \tfrac{1}{2 \gamma} \norm{ \vr - \vw }^2 +  \norm{\vr}_1 \right\}
\end{align*}
Since the cost function inside the $\argmin$ is non-smooth, we compute the sub-differential instead of the gradient. The optimality condition states that the vector $\vzero$ must be in the sub-differential set:
\begin{align*}
	\vzero &\in \tfrac{1}{\gamma} \left( \vr - \vw \right) + \partial_{\vr} \left(\norm{\vr}_1 \right) \\
	\implies 0 &\in r_i - w_i + \gamma \partial_{r_i} \left( | r_i |  \right) \qquad i = 1, \dots, N \\
	\implies 0 &\in r_i - w_i + \gamma \sign( r_i ) \qquad i = 1, \dots, N \\
	r_i &= \begin{cases}
	w_i - \gamma & \mbox{if } w_i > \gamma  \\
	w_i + \gamma & \mbox{if } w_i < -\gamma \\
	w_i & \mbox{otherwise}
	\end{cases}.
\end{align*}
Hence, the proximal operator for $k$ can be written in compact form as
\begin{align*}
	\prox_{\gamma k}(\vw) = \max \left(\vzero, \vw - \gamma  \right) - \max \left( \vzero, - \vw - \gamma  \right),
\end{align*}
where the operator $\max$ operates elementwise. 

\subsection{Optimality Conditions}
The optimality condition (aka stopping criterion) for primal-dual algorithm are based on two components. The first condition can be arrived at from the first order optimality condition
\begin{align*}
	\vzero \in \nabla h(\vmu) + \mA \vz \\
	\implies \quad \norm{ \vmu - \vy + \mA \vz} \leq \epsilon_{p},
\end{align*}
where $\epsilon_{p}$ is a tolerance criterion set by user (usually $10^{-6} - 10^{-4}$ works). The second condition can be derived from the progress of the iterates:
\begin{align*}
	\norm{ \vmu_{t+1} - \vmu_t} + \norm{\vz_{t+1} - \vz_t} \leq \epsilon_q,
\end{align*}
where $\epsilon_q$ is the tolerance (usually set to $10^{-8} - 10^{-6}$). These two stopping criteria are sufficient for practical purpose.  

\section{Numerical experiments - Binary tomography}
\begin{figure}[t]
\centering
\includegraphics[scale=.3]{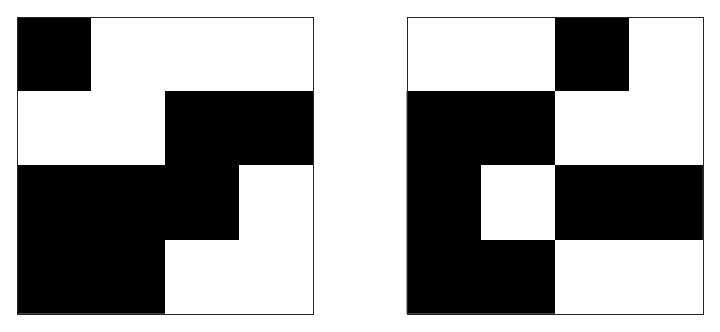}\\[2ex]
\includegraphics[scale=.5]{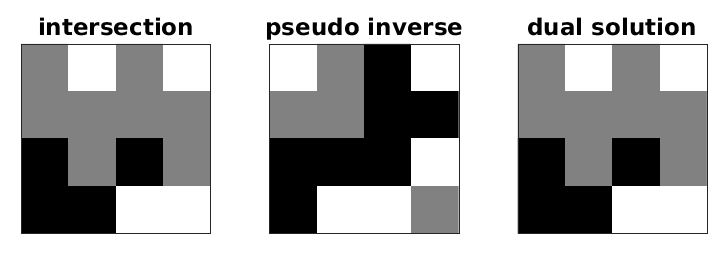}\\
\caption{Example for $n=4$ and $m = 3$ (using directons $(0,1)$, $(1,0)$ and $(1,1)$). Two images with the same projections are shown in the top row while the intersection and the results obtained by the pseudo-inverse and the dual problem are shown in the bottom row.}
\label{fig:examples}
\end{figure}

To illustrate the behaviour of the dual approach, we consider the simple setting of reconstructing an $n \times n$ image from
its sums along $m$ lattice directions (here restricted to the horizontal, vertical and two diagonal directions, so $m\in [2,4]$). For $m,n \geq 2$ the problem is known to be NP-hard.
For small $n$ we can simply enumerate all possible images, find all solutions in each case by a brute-force search and compare these to the solution obtained by the dual approach. To this end, we solved the resulting dual problem \eqref{eq:dual2} using the CVX package in Matlab \cite{grant2008cvx}. This yields an approximate solution, and we set elements in the numerically computed dual solution smaller than $10^{-9}$ to zero. We then compare the obtained primal solution, which has values ${-1,0,1}$ to the solution(s) of the binary tomography problem. From performing these computations for $n = 2,3,4$ and $m=2,3,4$ we conclude the following:

\begin{itemize}
\item If the problem has a unique solution then the dual approach retrieves it.
\item If the problem has multiple solutions then the dual approach retrieves the intersection of all solutions. The remaining pixels in the dual solution are undetermined (have value zero).
\end{itemize}
An example is shown in Figure~\ref{fig:examples}.

\begin{table}[t]
\caption{Summary of complete enumeration experiments.}
\centering
{\renewcommand{\arraystretch}{1.2}
\begin{tabular}{c | r| r | r | r}
\toprule
  & $\boldsymbol{n}$  & \textbf{total} & \textbf{unique} & \textbf{multiple} \\
\midrule
\multirow{3}{*}{\rotatebox[origin=c]{90}{$\boldsymbol{m=2}$}} & 2 &16&   14/14 &    2/2 \\
& 3  &512&  230/230 &   282/282 \\
& 4  &65536& 6902/6902 & 58541/58634$^*$ \\
\midrule
\multirow{3}{*}{\rotatebox[origin=c]{90}{$\boldsymbol{m=3}$}} & 2 &16&   16/16  &    0/0 \\
& 3 &512&  496/496  &   16/16\\
& 4 &65536&   54272/54272    &  10813/11264$^*$  \\
\midrule
\multirow{3}{*}{\rotatebox[origin=c]{90}{$\boldsymbol{m=4}$}} & 2 & 16&   16/16  &  0/0\\
& 3 &512&  512/512  & 0/0\\
& 4 &65536&  65024/65024     & 512/512   \\ \bottomrule
\end{tabular}
}
\label{tab:enumarateAll}
\end{table}

A summary of these results is presented in table \ref{tab:enumarateAll}. The table shows the number of cases with a unique solution where the dual approach gave the correct solution and, in case of multiple solutions, the number of cases where the dual approach correctly determined the intersection of all solutions). In a few instances with multiple solutions, CVX failed to provide an accurate solution (denoted with $^*$ in the table).

\begin{figure}[t]
\centering
\includegraphics[scale=.25]{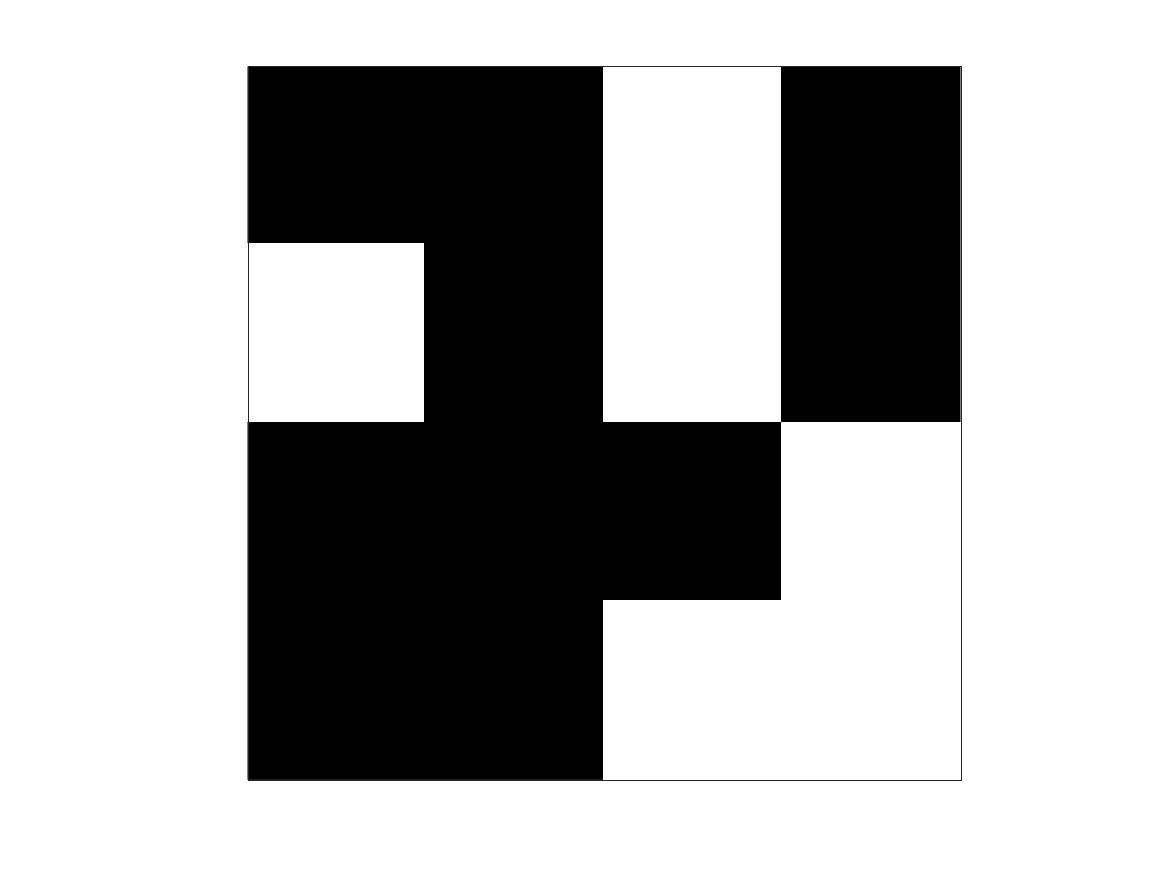}\\
\caption{Example of a $4\times 4$ binary image that is not \emph{h,v,d}-convex but does permit a unique solution.}
\label{fig:not_convex}
\end{figure}

Based on these experiments, we conjecture that there is a subclass of the described binary tomography problem that is not NP-hard. We should note that, as $n$ grows the number of cases that have a unique solution grows smaller unless $m$ grows accordingly. It has been established that binary images that are \emph{h,v,d}-convex\footnote{For \emph{h,v,d}-convexity one uses the usual definition of convexity of a set but considers only line segments in the horizontal, vertical and diagonal directions.} can be reconstructed from their horizontal, vertical and diagonal projections in polynomial time \cite{Barcucci1996,Barcucci2001}. However, we can construct images that are \emph{not} \emph{h,v,d}-convex but still permit a unique solution, see figure \ref{fig:not_convex}. Such images are also retrieved using our dual approach.

\section{Numerical experiments - X-ray tomography}
In this section, we present numerical results for limited-angle X-ray tomography on a few numerical phantoms and an experimental X-ray dataset. First, we describe the phantoms and the performance measures used to compare our proposed dual approach (abbreviated as DP) to several state-of-the-art iterative reconstruction techniques. We conclude this section with results on an experimental dataset.
All experiments are performed using Matlab in conjunction with the ASTRA toolbox \cite{van2015astra}.

\subsection{Phantoms}
\begin{figure*}[ht]
\centering
\subfloat[]{\includegraphics[width=1.2in]{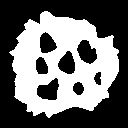}%
\label{fig:phantom1}}
\hfil
\subfloat[]{\includegraphics[width=1.2in]{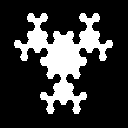}%
\label{fig:phantom2}}
\hfil
\subfloat[]{\includegraphics[width=1.2in]{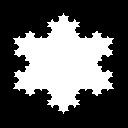}%
\label{fig:phantom3}}
\hfil
\subfloat[]{\includegraphics[width=1.2in]{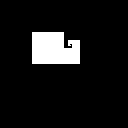}%
\label{fig:phantom4}}
\caption{Phantom images used in the simulation experiments. (a) Phantom 1, (b) Phantom 2, (c) Phantom 3, (d) Phantom 4.}
\label{fig:phantoms}
\end{figure*}

For the synthetic tests, we consider four phantoms shown in figure~\ref{fig:phantoms}. All the phantoms are binary images of size 128 $\times$ 128 pixels. The greylevels are $u_0 = 0$ and $u_1 = 1$. The detector has 128 pixels, and the distance between the adjacent detectors is the same as the pixel size of the phantoms. We consider a parallel beam geometry for the acquisition of the tomographic data in all the simulation experiments.

\subsection{Tests}
We perform three different tests to check the robustness of the proposed method. First, we consider the problem of sparse projection data. For all the phantoms, we first start with 45 projections at angles ranging from $0$ to $\pi$ and subsequently reduce the number of angles. This setup is also known as sparse sampling, where the aim is to reduce the scan time by decreasing the number of angles.

Next, we consider a limited angle scenario. Such situation usually arises in practice due to the limitations of the setup. For the test, we acquire projections in the range $[0,\theta_{\max}]$ for $\theta_{\text{max}} \in \{ 5 \pi/6, 2 \pi/3, 7\pi/12, \pi/2\}$. Reconstruction of limited angle data is known to lead to so-called streak artifacts in the reconstructed image. Strategies to mitigate these streak artifacts include the use of regularization method with some prior information. We will experiment how the discrete tomography can lead to the removal of these artifacts.

Finally, we test the performance of the proposed method in the presence of noise. We consider an additive Gaussian noise in these experiments. We measure the performance of our approach for tomographic data with an signal-to-noise ratio (SNR) of $\{10, 20, 30, 50 \}$ dB. 

To avoid inverse crime in all the test scenarios, we generate data using strip kernel and use Joseph kernel for modeling. 


\subsection{Comparison with other reconstruction methods}
There exist a vast amount of reconstruction methods for tomography. Here, consider the following three:

\begin{description}
\item[LSQR]: Least squares QR method described in \cite{paige1982lsqr}. We perform a total of 1000 iterations with a tolerance of $10^{-6}$. We segment the resulting reconstruction using Otsu's thresholding algorithm \cite{otsu1979threshold}.

\item[TV]: The total-variation method leads to an optimization problem described below:
\begin{align*}
	\underset{\vx \in \mathbb{R}^N}{\min} \quad \|\mA \vx - \vy\|^2 + \lambda \| \mathbf{D} \vx \|_1,
\end{align*}
where $\lambda$ is a corresponding regularization parameter, and $\mathbf{D}$ matrix captures the discrete gradient in both directions. We use the Chambolle-Pock method \cite{chambolle2011first} to solve the above optimization problem with non-negativity constraints on the pixel values. For each case, we perform iterations till the relative duality gap reach a tolerance value of $10^{-4}$. To avoid slow convergence, we scale the matrices $\mA$ and $\mathbf{D}$ to have unit matrix norm. The regularization parameter $\lambda$ is selected using Morozov's discrepancy principle using the correct noise level. We segment the TV-reconstructed image through Otsu's thresholding algorithm.

\item[DART]: We use the method described in \cite{batenburg2011dart} for DART on the above binary images. The grayvalues are taken to be the same as true grayvalues. We perform 20 ARM iterations initially before performing 40 DART iterations. In each DART iteration, we do 3 algebraic reconstruction iterations. We use the segmented image as a result of the DART iterations to perform the further analysis of the method.

\item[DP]: We solve the dual formulation \eqref{eq:dual2} with a smooth approximation of the $\ell_1$-norm (as discussed in section \ref{section:dualProblem}) using L-BFGS method \cite{liu1989limited} with a maximum of 500 iterations.
\end{description}

\subsection{Performance Measures}
In order to evaluate the performance of reconstruction methods, we use the following two criteria

\begin{description}

\item[RMS] The root-mean-square error
\begin{align*}
	\text{RMS} \triangleq \| \mA \vx^{\star} - \vy \|,
\end{align*}
measures how well the forward-projected reconstructed image matches the projection data. This measure is useful in practice, as it does not require knowledge of the ground truth. If the RMS value is close to the noise level of the data, the reconstruction is considered as a good reconstruction.

\item[JI] The Jaccard index $\mathrm{JI} \triangleq 1 - \sum_{i=1}^{N} (\alpha_i + \beta_i)/N$ measures the similarity between the reconstructed image ($\vx^\star$) and the ground truth ($\vx^{\text{true}}$) in a discrete sense. The parameters $\alpha$ and $\beta$ represent missing and over-estimated pixels respectively and given by
\begin{align*}
	\alpha_i &\triangleq \left( x_{i}^\star=u_0\right) \times \left(x_i^{\text{true}} = u_1 \right), \\
	\beta_i &\triangleq  \left( x_{i}^\star=u_1 \right) \times \left( x_i^{\text{true}} = u_0 \right).
\end{align*}
The blue and red dots denote the missing and the overestimated pixels in Figure~\ref{fig:LP:Exp1}, \ref{fig:LP:Exp2}, \ref{fig:LA:Exp1}, \ref{fig:LA:Exp2}, \ref{fig:NP}. If JI has high value (close to 1), the reconstruction is considered good. Although this measure can not generally be applied on real datasets, it is a handy measure to compare the various reconstruction methods on synthetic examples.

\end{description}
\subsection{Experimental data setup}
\begin{figure}[ht]
	\centering
	\includegraphics[width=0.5\columnwidth]{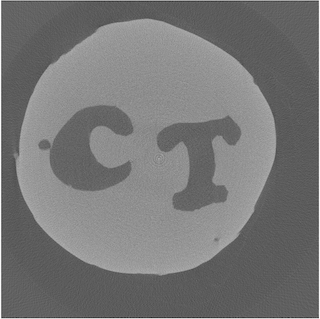}
	\caption{The high resolution ($2000\times 2000$ pixels) filtered back-projection reconstruction of the carved cheese from 360 projections from $0$ to $2 \pi$.}
	\label{fig:realDataFBP}
\end{figure}

We use the experimental X-ray projection data of a carved cheese slice\cite{bubba2017tomographic}. Figure~\ref{fig:realDataFBP} shows a high-resolution filtered back-projection reconstruction of the data. The cheese contains the letters C and T and the object is (approximately) binary with two grey levels corresponding to calcium-containing organic compounds of cheese and air. The dataset consists of projection data with three different resolutions ($128 \times 128$, $256 \times 256$, $512 \times 512$) and the corresponding projection matrix modelling the linear operation of X-ray transform.
We perform two sets of experiments: (1) Sparse sampling with 15 angles ranging from $0$ to $2 \pi$ and (2) limited-angle using 15 projections from $0$ to $\pi/2$.

\subsection{Sparse projections test}
\begin{figure}
	\centering
	\subfloat[LSQR]{\includegraphics[width=0.7in]{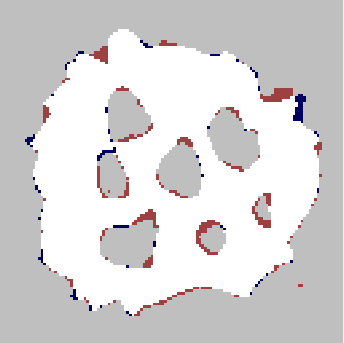}%
	\label{fig:LP:m1:LSQR}}
	\hfil
	\subfloat[TV]{\includegraphics[width=0.7in]{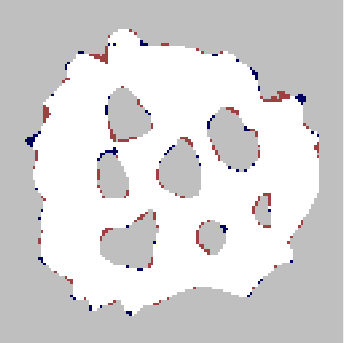}%
	\label{fig:LP:m1:TV}}
	\hfil
	\subfloat[DART]{\includegraphics[width=0.7in]{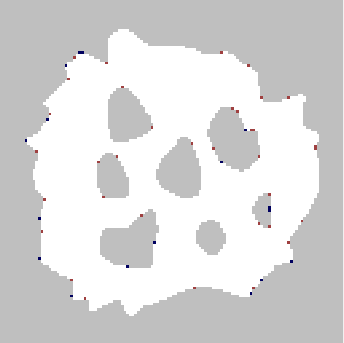}%
	\label{fig:LP:m1:DART}}
	\hfil
	\subfloat[DP]{\includegraphics[width=0.7in]{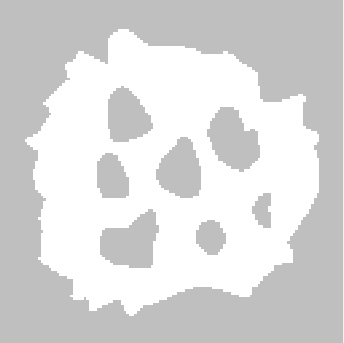}%
	\label{fig:LP:m1:DP}}
	\caption{Limited projection test I for Phantom 1. Performance of various reconstruction methods with 10 projection angles from 0 to $\pi/2$.} 
	\label{fig:LP:Exp1}
\end{figure}

Figure~\ref{fig:LP:Exp1} presents the reconstruction results from various methods for phantoms 1. The tomographic data are generated for ten equidistant projection angles from $0$ to $\pi/2$. The reconstruction results show the difference between the reconstructed image and the ground truth.  It is evident that, compared to the other methods, the proposed method reconstructions are very close to the ground truth. The results from LSQR are the worst as it does not incorporate any prior information about the model. The TV method also leads to artifacts as it includes the partial information about the model. The DART and DP are very close to each other. For all the phantoms, we tabulate the data misfit (RMS) and Jaccard index (JI) in Table~\ref{table:LP}.

\begin{figure}
	\centering
	\subfloat[45]{\includegraphics[width=0.7in]{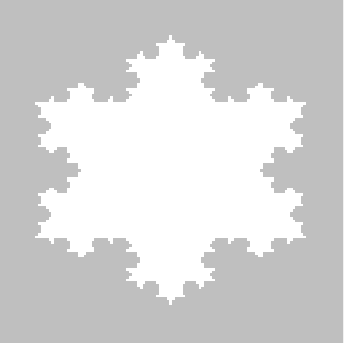}%
	\label{fig:LP:m3:DP:45}}
	\hfil
	\subfloat[20]{\includegraphics[width=0.7in]{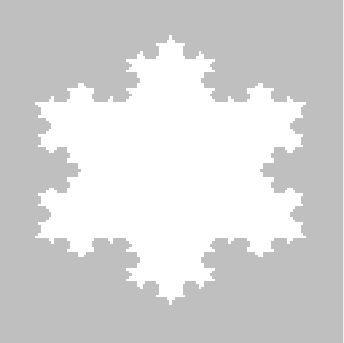}%
	\label{fig:LP:m3:DP:20}}
	\hfil
	\subfloat[10]{\includegraphics[width=0.7in]{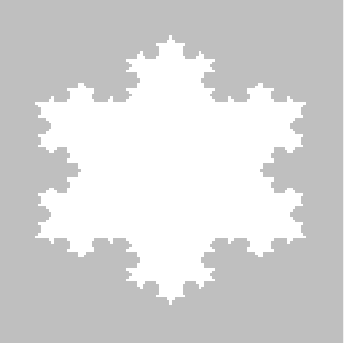}%
	\label{fig:LP:m3:DP:10}}
	\hfil
	\subfloat[5]{\includegraphics[width=0.7in]{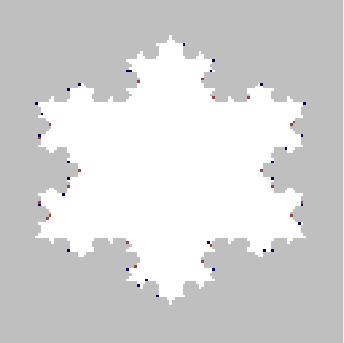}%
	\label{fig:LP:m3:DP:5}}
	\caption{Limited projection test II for Phantom 3. Performance of proposed method vs number of projections.} 
	\label{fig:LP:Exp2}
\end{figure}

We also show reconstructions with the proposed approach for a varying number of projection angles in Figure~\ref{fig:LP:Exp2}. We note that the problem becomes harder to solve as the number of projection angles gets smaller. Hence, we may also expect the reconstruction to become poorer. We see that the proposed approach can reconstruct almost correctly with as few as ten projection angles.

\begin{table*}
	\caption{Limited projection test performance measures}
	\centering
	{\setlength{\tabcolsep}{0.7em}
	\begin{tabular}{c|c|cc|cc|cc|cc}
		\toprule
		\multirow{2}{*}{\rotatebox[origin=c]{90}{\textbf{Test}}} & \multirow{2}{*}{ \rotatebox[origin=c]{90}{\textbf{Phantom}} } & \multicolumn{2}{c|}{\textbf{LSQR}} & \multicolumn{2}{c|}{\textbf{TV}} & \multicolumn{2}{c|}{\textbf{DART}} & \multicolumn{2}{c}{\textbf{DP}} \\[3.2ex]
		& & \rotatebox[origin=c]{90}{RMS} & \rotatebox[origin=c]{90}{JI} & \rotatebox[origin=c]{90}{RMS} & \rotatebox[origin=c]{90}{JI} & \rotatebox[origin=c]{90}{RMS} & \rotatebox[origin=c]{90}{JI} & \rotatebox[origin=c]{90}{RMS} & \rotatebox[origin=c]{90}{JI} \\ \midrule
		\multirow{4}{*}{\rotatebox[origin=c]{90}{\textbf{45}}} 	& P1 & 31.4 & 99.7 & 19.2 & 99.9 & 31.5 & 99.7 & 17.6 & 100 \\
							& P2 & 26.2 & 99.6 & 12 & 100 & 36.4 & 99 & 12 & 100 \\
							& P3 & 48.1 & 99 & 24.5 & 99.8 & 52.4 & 98.8 & 16.4 & 100 \\
							& P4 & 11.6 & 99.8 & 5  & 100 & 13.6 & 99.6 & 5  & 100 \\ \midrule
		\multirow{4}{*}{\rotatebox[origin=c]{90}{\textbf{20}}} & P1 & 54.3 & 97.8 & 34.5 & 98.9 & 24.1 & 99.5 & 12 & 100 \\
							& P2 & 60.5 & 95.8 & 12.8 & 99.9 & 17 & 99.6 & 8.9  & 100 \\
							& P3 & 54.2 & 97.6 & 27 & 99.4 & 40.2 & 98.9 & 11.1 & 100 \\
							& P4 & 20.8 & 99.3 & 3.4  & 100 & 6.8  & 99.8 & 3.4  &	 100 \\ \midrule
		\multirow{4}{*}{\rotatebox[origin=c]{90}{\textbf{10}}}	& P1 & 122.3 & 93.1 & 62.4 & 96.2 & 22.4 & 99.1 & 8.3  & 99.9 \\
							& P2 & 254.4 & 76 & 81 & 91.1 & 17 & 99 & 11.2 & 99.7 \\
							& P3 & 73.8 & 94.9 & 38.6 & 98 & 28.7 & 99 & 7.4 & 100 \\
							& P4 & 42.7 & 97 & 5.4 & 99.9 & 5.9 & 99.7 & 2.4 & 00 \\ \midrule
		\multirow{4}{*}{\rotatebox[origin=c]{90}{\textbf{5}}} 	& P1 & 269.7 & 82.4 & 77.5 & 87 & 23.7 & 97.3 & 52.8 & 90.7 \\
							& P2 & 370.8 & 63.2 & 158.5 & 71.3 & 38.8 & 76.1 & 72.8 & 73.9 \\
							& P3 & 80 & 90.9 & 64.9 & 93.4 & 22.5 & 98.5 & 21.5 & 97.6 \\
							& P4 & 79.7 & 86.4 & 22.9 & 95.8 & 4.7 & 99.6 & 1.7 & 100 \\ \bottomrule
	\end{tabular}
	}
	\label{table:LP}
\end{table*}

\subsection{Limited angle test}
\begin{figure}
	\centering
	\subfloat[LSQR]{\includegraphics[width=0.7in]{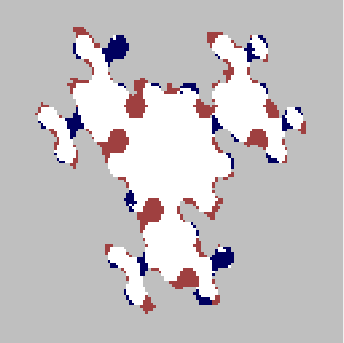}%
	\label{fig:LA:m2:LSQR}}
	\hfil
	\subfloat[TV]{\includegraphics[width=0.7in]{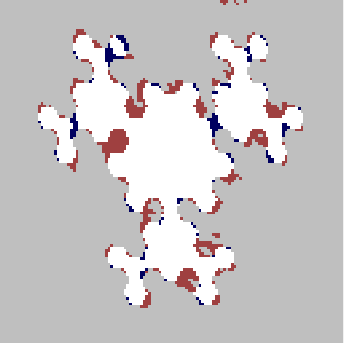}%
	\label{fig:LA:m2:TV}}
	\hfil
	\subfloat[DART]{\includegraphics[width=0.7in]{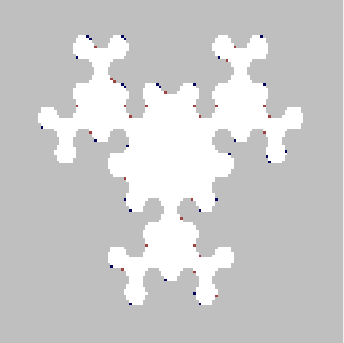}%
	\label{fig:LA:m2:DART}}
	\hfil
	\subfloat[DP]{\includegraphics[width=0.7in]{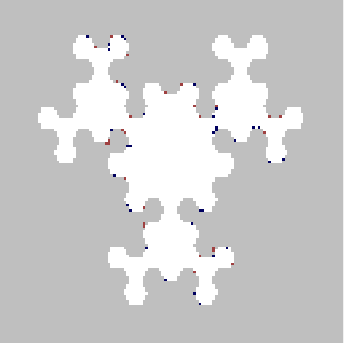}%
	\label{fig:LA:m2:DP}}
	\caption{Limited angle test I for Phantom 2. Performance of various reconstruction methods with 10 projection angles from 0 to $\tfrac{\pi}{2}$.} 
	\label{fig:LA:Exp1}
\end{figure}

Figure~\ref{fig:LA:Exp1} shows the results of phantom 2 with various reconstruction methods for limited angle tomography ( 10 equispaced angles in the range $0$ to $\pi/2$). It is visible that the reconstructions from DART and the proposed method are very close to true images of the phantoms. The values for data misfit and Jaccard index for all the tests with each of the synthetic phantoms are tabulated in Table~\ref{table:LA}.

\begin{figure}
	\centering
	\subfloat[$5 \pi/6$]{\includegraphics[width=0.7in]{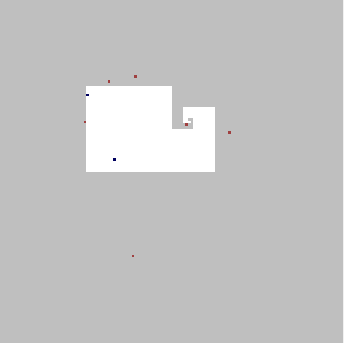}%
	\label{fig:LA:m4:DP:135}}
	\hfil
	\subfloat[$2 \pi /3$]{\includegraphics[width=0.7in]{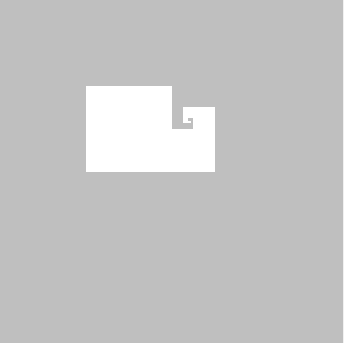}%
	\label{fig:LA:m4:DP:120}}
	\hfil
	\subfloat[$7 \pi/12$]{\includegraphics[width=0.7in]{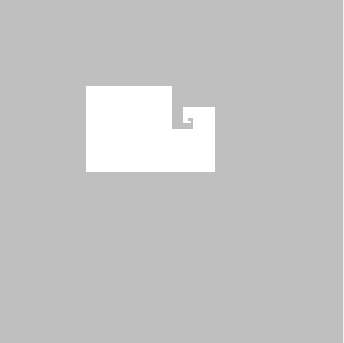}%
	\label{fig:LA:m4:DP:105}}
	\hfil
	\subfloat[$\pi/2$]{\includegraphics[width=0.7in]{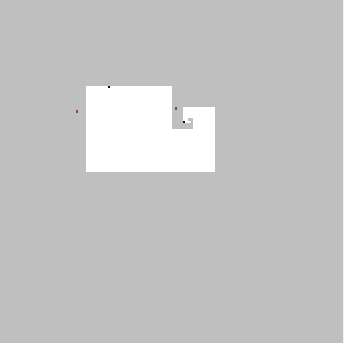}%
	\label{fig:LA:m4:DP:90}}
	\caption{Limited angle test II for Phantom 4. Performance of proposed method vs maximum angle.} 
	\label{fig:LA:Exp2}
\end{figure}

We also look at how the reconstructions with the proposed method varies with limiting the angle (see Figure~\ref{fig:LA:Exp2}). As the angle gets limited, the reconstruction problem gets difficult. The proposed method can reconstruct almost perfectly with angle limited to $\pi/2$.

\begin{table*}
	\caption{Limited angle test performance measures}
	\centering
	{\setlength{\tabcolsep}{0.7em}
	\begin{tabular}{c|c|cc|cc|cc|cc}
		\toprule
		\multirow{2}{*}{ \rotatebox[origin=c]{90}{\textbf{Test}}} & \multirow{2}{*}{\rotatebox[origin=c]{90}{\textbf{Phantom}}} & \multicolumn{2}{c|}{\textbf{LSQR}} & \multicolumn{2}{c|}{\textbf{TV}} & \multicolumn{2}{c|}{\textbf{DART}} & \multicolumn{2}{c}{\textbf{DP}} \\[3.2ex]
		& & \rotatebox[origin=c]{90}{RMS} & \rotatebox[origin=c]{90}{JI} & \rotatebox[origin=c]{90}{RMS} & \rotatebox[origin=c]{90}{JI} & \rotatebox[origin=c]{90}{RMS} & \rotatebox[origin=c]{90}{JI} & \rotatebox[origin=c]{90}{RMS} & \rotatebox[origin=c]{90}{JI} \\ \midrule
		\multirow{4}{*}{\rotatebox[origin=c]{90}{$\boldsymbol{5\pi/6}$}} & P1 & 158.3 & 91.6 & 101.9 & 93.9 & 23.6 & 99.1 & 5.4 & 100 \\
							& P2 & 215.9 & 79.5 & 79.6 & 91.3 & 16.4 & 99.2 & 5 & 100 \\
							& P3 & 82.4 & 94.5 & 53.3 & 96.7 & 28.5 & 99 & 5.3 & 100 \\
							& P4 & 54.9 & 95.6 & 4.7 & 99.9 & 5.3 & 99.7 & 9.5 & 99.2 \\ \midrule
		\multirow{4}{*}{\rotatebox[origin=c]{90}{$\boldsymbol{2\pi/3}$}}	& P1 & 199.7 & 88.4 & 141.7 & 91.5 & 18.7 & 99.3 & 6.7 & 100 \\
							& P2 & 189.3 & 77.8 & 146.5 & 83 & 22 & 98 & 13.3 & 99.3 \\
							& P3 & 96.2 & 92.9 & 70.5 & 95 & 27.4 & 98.6 & 3.4 & 100 \\
							& P4 & 68.2 & 91.7 & 9.1 & 99.3 & 7.1 & 99.5 & 0.8 & 100 \\ \midrule
		\multirow{4}{*}{\rotatebox[origin=c]{90}{$\boldsymbol{7\pi/12}$}} & P1 & 213.2 & 87 & 164.6 & 89.7 & 20.1 & 99.2 & 7.7 & 99.9 \\
							& P2 & 231.8 & 76.4 & 182.1 & 81.2 & 21.7 & 98 & 17 & 98.5 \\
							& P3 & 105.5 & 92.3 & 79.9 & 94.3 & 28.3 & 98.5 & 3.7 & 100 \\
							& P4 & 84.5 & 89.7 & 8.1 & 99.4 & 7.6 & 99.5 & 0.9 & 100 \\ \midrule
		\multirow{4}{*}{\rotatebox[origin=c]{90}{$\boldsymbol{\pi/2}$}}  & P1 & 258.6 & 84.9 & 293 & 85.2 & 19.1 & 99.3 & 18 & 99.2 \\
							& P2 & 205.3 & 75.3 & 192.5 & 80.9 & 22.7 & 98.2 & 18.2 & 98.5 \\
							& P3 & 127.1 & 90.5 & 99.2 & 93.5 & 31 & 98.3 & 4 & 100 \\
							& P4 & 128.2 & 82.9 & 27.1 & 96.7 & 7.8 & 0.99.5 & 7.2 & 99.5 \\ \bottomrule
	\end{tabular}
	}
	\label{table:LA}
\end{table*}

\subsection{Noisy projection test}
\begin{figure*}
	\centering
	\subfloat[50 dB]{\includegraphics[width=1.4in]{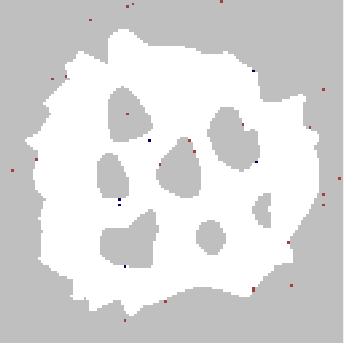}%
	\label{fig:NP:m1:DP:50}}
	\hfil
	\subfloat[30 dB]{\includegraphics[width=1.4in]{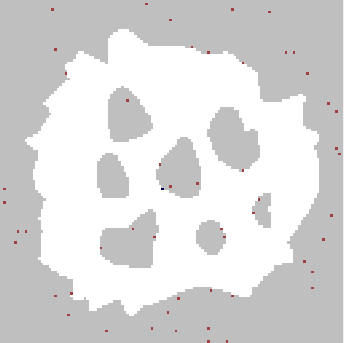}%
	\label{fig:NP:m1:DP:40}}
	\hfil
	\subfloat[10 dB]{\includegraphics[width=1.4in]{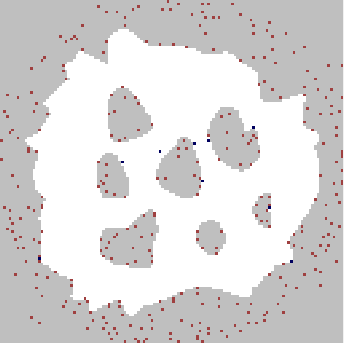}%
	\label{fig:NP:m1:DP:30}}
	\hfil
	\subfloat[5 dB]{\includegraphics[width=1.4in]{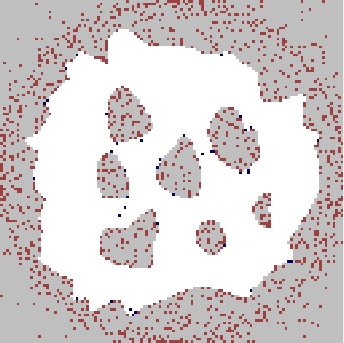}%
	\label{fig:NP:m1:DP:20}} \\
	\hfil
	\subfloat[50 dB]{\includegraphics[width=1.4in]{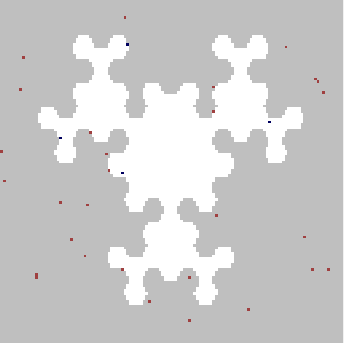}%
	\label{fig:NP:m2:DP:50}}
	\hfil
	\subfloat[30 dB]{\includegraphics[width=1.4in]{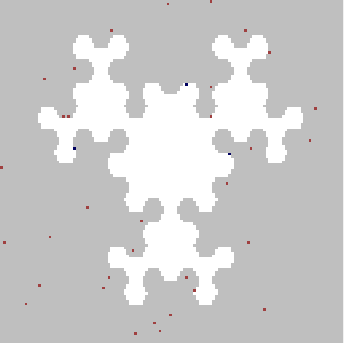}%
	\label{fig:NP:m2:DP:40}}
	\hfil
	\subfloat[10 dB]{\includegraphics[width=1.4in]{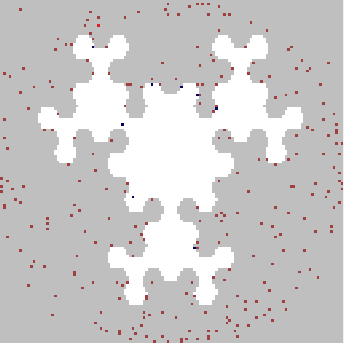}%
	\label{fig:NP:m2:DP:30}}
	\hfil
	\subfloat[5 dB]{\includegraphics[width=1.4in]{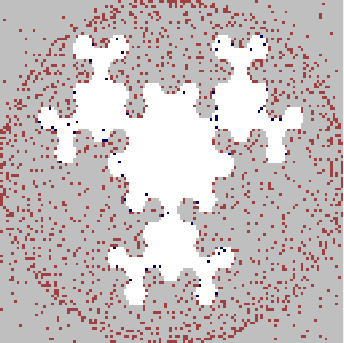}%
	\label{fig:NP:m2:DP:20}}
	\caption{Noisy Projection test on phantoms 1 and 2. Performance of proposed method vs signal-to-noise ratio.} 
	\label{fig:NP}
\end{figure*}

This test aims to check the sensitivity of the proposed method to noise in the data. We perform four experiments with varying levels of Poisson noise in the data. In particular, we use incident photon counts $I_0 = \left\lbrace 10^6, 10^4, 10^3, 10^2 \right\rbrace$ which leads to an approximate signal-to-noise-ratio of $\{50, 30, 10, 5\}$dB respectively. Figure~\ref{fig:NP} shows the results on phantoms 1 and 2 for increasing noise level. We see that the reconstruction is stable against a moderate amount of noise and degrades gradually as the noise level increases.

\subsection{Real data test}
We look at the results of reconstructions from the proposed method for two sets of experiments at various resolutions and compare them with the reconstructions from LSQR and TV. Since the ground truth image is not available, we compare these reconstructions visually.

\begin{figure}
	\centering
	\includegraphics[width=0.75\columnwidth]{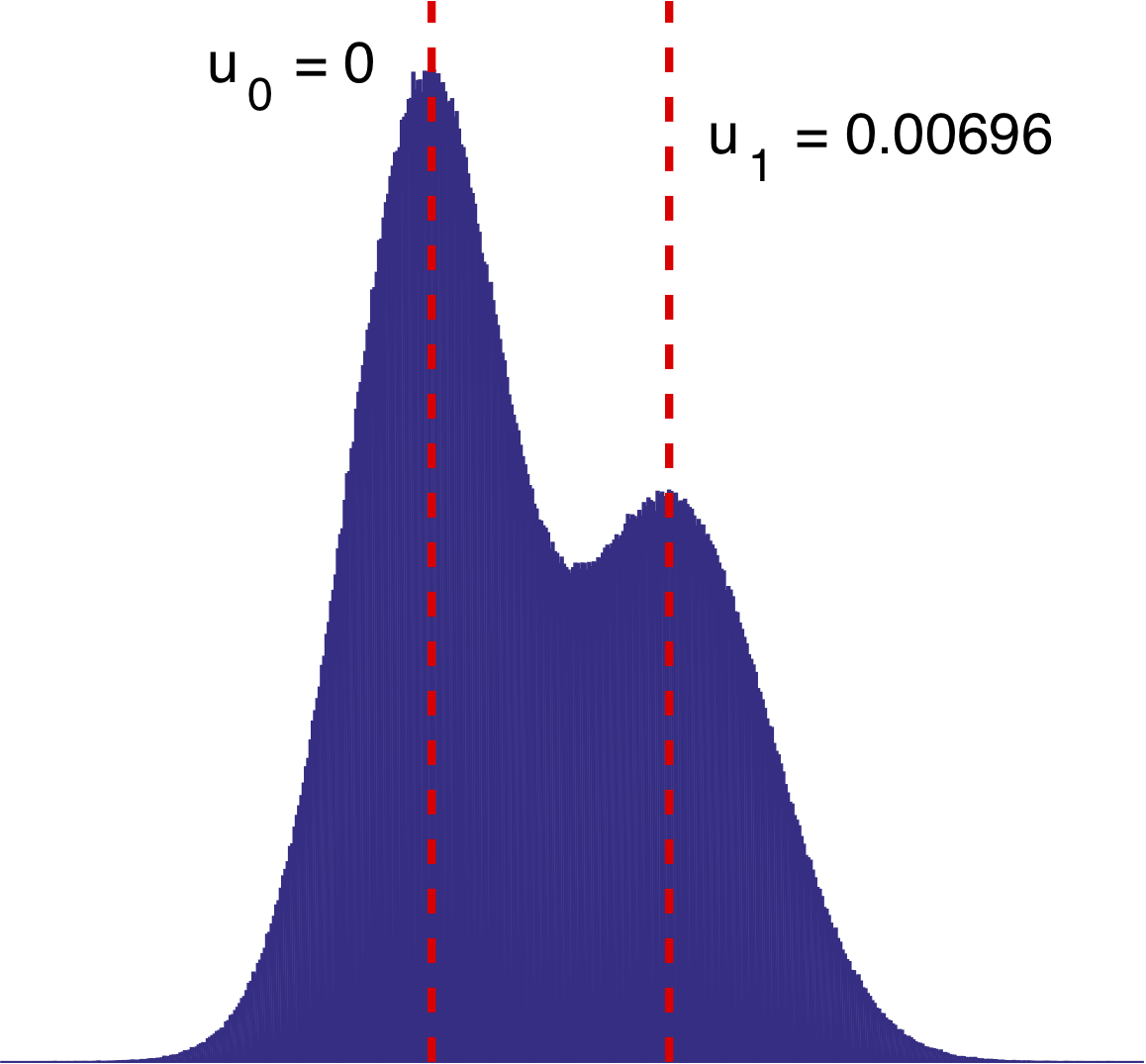}
	\caption{Histogram of filtered backprojection image of the carved cheese.}
	\label{fig:hist}
\end{figure}

In order to apply DP, we first need to estimate the grey values of the object. The object, a thin slice of cheese, consists of two materials; the organic compound of the cheese, which is we assume to be homogeneous, and air. For air, the grey value is zero. We estimate the grey value of the organic compound of cheese from the histogram of an FBP reconstruction provided with the data. Figure \ref{fig:hist} represents the histogram. We obtain a value of 0.00696 for this compound.

\begin{figure}
	\centering
	\subfloat[LSQR]{\begin{tabular}{c}
  	\includegraphics[width=0.9in]{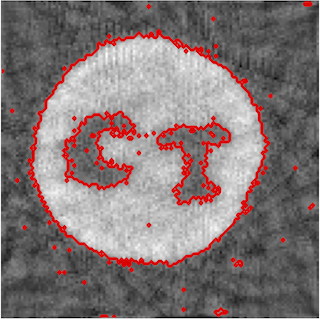} \\
	\end{tabular}
	\label{fig:RD:F128:LS}}
	\hfil
	\subfloat[TV]{\begin{tabular}{c}
  	\includegraphics[width=0.9in]{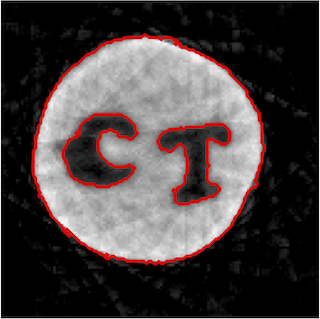} \\
	\end{tabular}
	\label{fig:RD:F128:TV}}
	\hfil
	\subfloat[DP]{\begin{tabular}{c}
  	\includegraphics[width=0.9in]{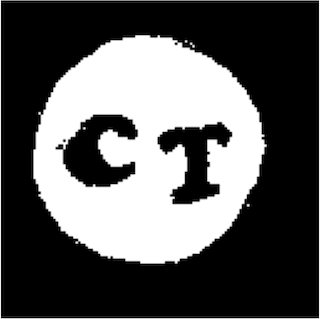} \\
	\end{tabular}
	\label{fig:RD:F128:DP}}
	\hfil
	\subfloat[LSQR]{\begin{tabular}{c}
	\includegraphics[width=0.9in]{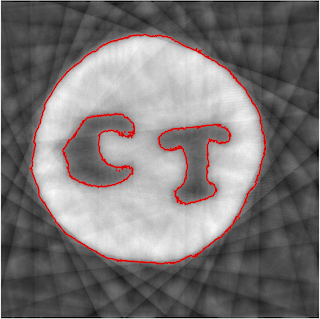} \\
	\end{tabular}
	\label{fig:RD:F512:LS}}
	\hfil
	\subfloat[TV]{\begin{tabular}{c}
	\includegraphics[width=0.9in]{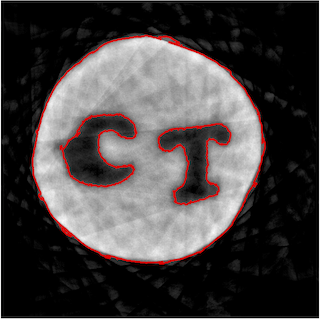} \\
	\end{tabular}
	\label{fig:RD:F512:TV}}
	\hfil
	\subfloat[DP]{\begin{tabular}{c}
	\includegraphics[width=0.9in]{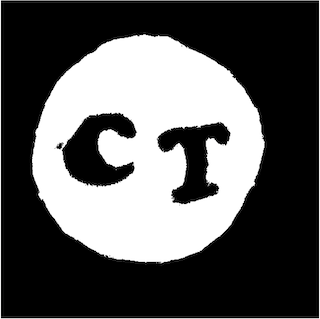} \\
	\end{tabular}
	\label{fig:RD:F512:DP}}
	\caption{Real Data Test I - Sparse projection tomography. Performance of various methods with different resolutions. Top row corresponds to $128 \times 128$ pixels. Bottom row corresponds to $512 \times 512$ pixels. Figure below each image denote the histogram. The red contours represent the thresholded image.}
	\label{fig:realData:spAngles}
\end{figure}

We first consider the reconstructions from sparse angular sampling. We have a tomographic data from 15 projections spanning from $0$ to $2 \pi$. The tests are performed on two different resolutions: $128 \times 128$, and $512 \times 512$. Figure~\ref{fig:realData:spAngles} presents the results of the reconstructions with LSQR, TV and DP for these resolutions. The DP reconstruction is discrete and correctly identifies the letters C and T with also a little hole at the left side of C. Although LSQR reconstruction is poor for $128\times128$, it improves with the resolution. We still see the mild streak artifacts in these reconstructions. The TV reconstruction removes these streak artifacts but fails to identify the homogeneous cheese slice correctly.

\begin{figure}[!htb]
	\centering
	\subfloat[LSQR]{\begin{tabular}{c}
	\includegraphics[width=0.9in]{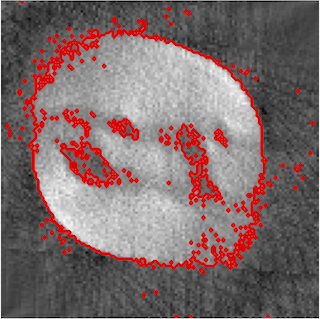} \\
	\end{tabular}
	\label{fig:RD:L128:LS}}
	\hfil
	\subfloat[TV]{\begin{tabular}{c}
	\includegraphics[width=0.9in]{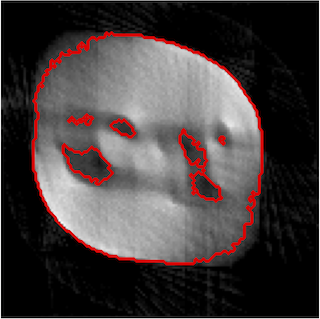} \\
	\end{tabular}
	\label{fig:RD:L128:TV}}
	\hfil
	\subfloat[DP]{\begin{tabular}{c}
	\includegraphics[width=0.9in]{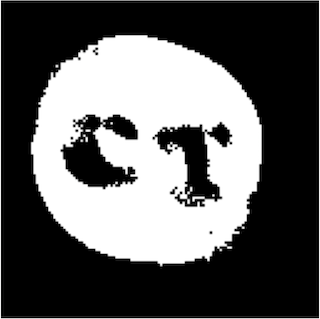} \\
	\end{tabular}
	\label{fig:RD:L128:DP}}
	\hfil
	\subfloat[LSQR]{\begin{tabular}{c}
	\includegraphics[width=0.9in]{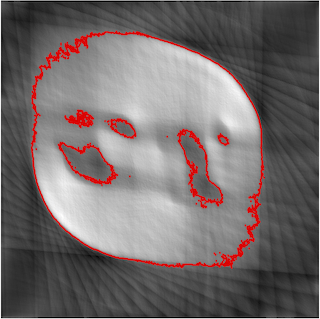} \\
	\end{tabular}
	\label{fig:RD:L512:LS}}
	\hfil
	\subfloat[TV]{\begin{tabular}{c}
	\includegraphics[width=0.9in]{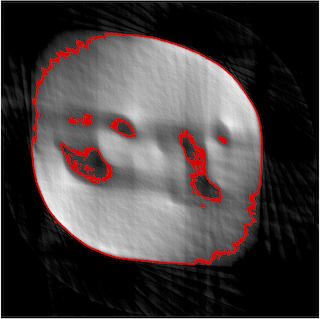} \\
	\end{tabular}
	\label{fig:RD:L512:TV}}
	\hfil
	\subfloat[DP]{\begin{tabular}{c}
	\includegraphics[width=0.9in]{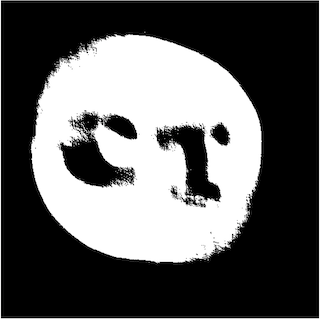} \\
	\end{tabular}
	\label{fig:RD:L512:DP}}
	\caption{Real Data Test II - Limited angle tomography. Performance of various methods with different resolutions. Top row corresponds to $128 \times 128$ pixels. Bottom row corresponds to $512 \times 512$ pixels. Figure below each image denote the histogram. The red contours represent the thresholded image.}
	\label{fig:realData:limited}
\end{figure}

In the second test, we limit the projection angles to $0 - \pi/2$. Figure~\ref{fig:realData:limited} shows the results of the reconstructions from LSQR, TV, and DP for two different resolutions. We see that the reconstructions improve with increment in the resolution. LSQR reconstructions have severe streak artifacts, which are the characteristics of the limited data tomography. TV and DP reconstructions do not possess these artifacts. TV reconstruction can capture the shape of the cheese, but it blurs out the carved parts C and T. DP reconstructs the shape of cheese quite accurately and has C and T are also identified.

\section{Conclusion}
We presented a novel convex formulation for binary tomography. The problem is primarily a generalized LASSO problem that can be solved efficiently when the system matrix has full row rank or full column rank. Solving the dual problem is not guaranteed to give the optimal solution, but can at least be used to construct a feasible solution. In a complete enumeration of small binary test cases (images of $n\times n$ pixels for $n = 2, 3, 4$) we observed that if the problem has a unique solution, then the proposed dual approach finds it. In case the problem has multiple solutions, the dual approach finds the part that is common in all solutions. Based on these experiments we conjecture that this holds in the general case (beyond the small test images). Of course, verifying beforehand if the problem has a unique solution may not be possible.

We test the proposed method on numerical phantoms and real data, showing that the method compares favourably to some of the state-of-the-art reconstruction techniques (Total Variation, DART). The proposed method is also reasonably stable against a moderate amount of noise.

We currently assume the grey levels are known apriori. Extension to multiple (i.e., more than 2) unknown grey levels is possible in the same framework but will be left for future work. To make the method more robust against noise additional regularization may be added.

\appendices

\numberwithin{equation}{section}
\section{Proofs}

\subsection{Proposition \ref{thm1}}
\begin{proof}
In \eqref{eq:DualFunction}, the $g_1(\boldsymbol{\nu})$ has a closed-form expression for general $\mA$. To see this, let us first denote
\begin{equation}
	f(\vx,\boldsymbol{\nu}) \triangleq \tfrac{1}{2} \| \vy - \mA \vx \|^2 + \boldsymbol{\nu}^T \vx.
	\label{eq:functionForg1}
\end{equation}
We are interested in the infimum value of this function with repsect to $\vx$. To obtain this, we set the gradient of $f$ with respect to $\vx$ to zero
\begin{align*}
	\nabla_{\vx} {f} = \mA^T \!\left( \mA \vx - \vy \right) + \boldsymbol{\nu} = 0.
\end{align*}
Since $\mA$ is a general matrix, it may be rank-deficient. Hence, the optimal value $\vx^\star$ only exists if $\boldsymbol{\nu}$ is in the range of $\mA^T$ (same as the row space of $\mA$) and it is given by
\[
	\vx^\star = \left( \mA^T\! \mA \right)^\dagger \left( \mA^T \vy - \boldsymbol{\nu} \right),
\]
where $\dagger$ denotes the Moore-Penrose pseudo-inverse of the matrix.
Substituting this value in \eqref{eq:functionForg1}, we get the following:
\begin{align*}
	g_1(\boldsymbol{\nu}) &=
	\inf_{\vx} f( \vx ,\boldsymbol{\nu}) \\
	 &= \begin{cases}
	f(\vx^\star ,\boldsymbol{\nu}) & \boldsymbol{\nu} \in \mathcal{R}_{\mA} \\
	-\infty & \text{otherwise}
	\end{cases} \\
	&= \begin{cases}
	-\tfrac{1}{2} \| \mathbf{\nu} - \mA^T \vy \|_{\left(\mA^T\! \mA\right)^{\dagger}} + \tfrac{1}{2} \vy^T \vy & \boldsymbol{\nu} \in \mathcal{R}_{\mA} \\
	-\infty & \text{otherwise},
	\end{cases}
\end{align*}
where $\mathcal{R}_{\mA}$ denotes the row-space of $\mA$.

Now we return to the dual objective in equation~\eqref{eq:DualFunction}. Substituting the explicit forms for $g_1(\boldsymbol{\nu})$ from above and $g_2(\boldsymbol{\nu})$ from equation~\eqref{eq:Explicitg2}, we get the expression for the dual objective:
\begin{align*}
g(\boldsymbol{\nu}) =
\begin{cases}
-\tfrac{1}{2} \| \boldsymbol{\nu} - \mA^T \vy\|_{\left( \mA^T \!\mA \right)^{\dagger}}^2 - \| \boldsymbol{\nu} \|_1 + \tfrac{1}{2} \vy^T \vy & \boldsymbol{\nu} \in \mathcal{R}_{\mA}, \\
-\infty & \text{otherwise}.
\end{cases}
\end{align*}
The above dual objective leads to the following maximization problem with respect to the dual variable $\boldsymbol{\nu}$
\begin{align*}
	\underset{\boldsymbol{\nu} \in \mathbb{R}^N}{\max} \quad g(\boldsymbol{\nu}).
\end{align*}
As we are only interested in the maximum value of the dual objective, the space of $\boldsymbol{\nu}$ can be constrained to the range of $\mA^T$. This is valid as the dual objective is $- \infty$ for the $\boldsymbol{\nu}$ outside the range of $\mA^T$. Hence, the maximization problem reduced to the following minimization problem:
\begin{align*}
	\underset{\boldsymbol{\nu} \in \mathcal{R}_{\mA}}{\min} \quad \tfrac{1}{2} \| \boldsymbol{\nu} - \mA^T \vy\|_{\left( \mA^T \!\mA \right)^{\dagger}}^2 + \| \boldsymbol{\nu} \|_1.
\end{align*}
\end{proof}

\subsection{Corollary \ref{Corollary:1st}}
\begin{proof}
Since the search space for the dual variable $\boldsymbol{\nu}$ is constrained to the range of $\mA^T$, we can express this variable as $\mA^T \vmu$, where $\vmu \in \mathbb{R}^m$. Substituting $\boldsymbol{\nu} = \mA^T \vmu$ in \eqref{eq:dual1}, we get
\begin{align}
	\underset{\vmu \in \mathbb{R}^m}{\min} \quad  \tfrac{1}{2} \| \mA^T \left( \vmu - \vy \right) \|_{\left( \mA^T \!\mA \right)^{\dagger}}^2 + \| \mA^T \vmu \|_1.
	\label{eq:initialFormwithMu}
\end{align}
Using the identities $(\mA^T\mA)^\dagger \mA^T = \mA^\dagger$ and $\mA = \mA\mA^\dagger \mA$ \cite{barnett1990matrices}, we can re-write the weighted norm $\|\mA^T\vr\|_{\left( \mA^T \!\mA \right)^{\dagger}}^2$ as $\|AA^\dagger\vr\|$.
The dual problem \eqref{eq:initialFormwithMu} now reads
\begin{align*}
	\underset{\vmu \in \mathbb{R}^m}{\min} \quad  \tfrac{1}{2} \| \mA \mA^{\dagger} \left( \vmu - \vy \right) \|^2 + \| \mA^T \vmu \|_1.
\end{align*}
The optimal solution to the above problem is denoted by $\vmu^\star$. Correspondingly, the primal solution $\vx^\star$ related to the dual optimal $\mathbf{\mu}^\star$ is
\[
	\vx^\star = \sign( \boldsymbol{\phi^\star}) = \sign \left( \boldsymbol{\nu}^\star \right) = \sign \left( \mA^T \boldsymbol{\nu}^\star \right).
\]
\end{proof}

\subsection{Proposition \ref{Corollary:2nd}}
\begin{proof}
The primal problem for binary tomography problem with grey levels $u_0 < u_1$ can be stated as:
\begin{align*}
	\underset{\vx \, , \, \vphi }{\min} \quad & \tfrac{1}{2} \| \mA \vx - \vy \|^2, \\
	\mbox{subject to} \quad & \vx = u_0 \mathbf{1} + \left( u_1 - u_0 \right) H(\vphi),
\end{align*}
where $H(\cdot)$ denotes the Heaviside function and $\vphi$ is an auxiliary variable. Such problem admits a Lagrangian
\[
	\mathcal{L}(\vx, \vphi, \boldsymbol{\nu}) = \tfrac{1}{2} \| \mA \vx - \vy \|^2 + \boldsymbol{\nu}^T \left( \vx - u_0 \mathbf{1} - \left( u_1 - u_0 \right) H(\vphi) \right) ,
\]
where $\boldsymbol{\nu} \in \mathbb{R}^N$ is a Lagrangian multiplier (also known as dual variable) corresponding to the equality constraint. This gives rise to a dual function
\begin{align*}
	g(\boldsymbol{\nu}) &= \underbrace{\inf_{\vx} \left\lbrace \tfrac{1}{2} \| \mA \vx - \vy \|^2 + \boldsymbol{\nu}^T \vx \right\rbrace}_{g_1(\boldsymbol{\nu})} \\
	& \quad + \underbrace{\inf_{\vphi} \left\lbrace - \left(u_1 - u_0 \right) \boldsymbol{\nu}^T H(\vphi) \right\rbrace}_{g_2(\boldsymbol{\nu})} - u_0 \boldsymbol{\nu}^T \mathbf{1}.
\end{align*}
Since we already know $g_1(\boldsymbol{\nu})$ (refer to equation~\eqref{eq:DT-P:LS:g1}), we require the explicit form for $g_2(\boldsymbol{\nu})$. For its computation, we use the componentwise property of the Heaviside function to separate the infimum.
\begin{align*}
	g_2(\boldsymbol{\nu}) &= \sum_{i=1}^{N} \inf_{\phi_i} \left\lbrace - \left(u_1 - u_0 \right) \nu_i H(\phi_i)\right\rbrace \\
	&= \sum_{i=1}^{N} \sup_{\phi_i} \left\lbrace  \left(u_1 - u_0 \right) \nu_i H(\phi_i) \right\rbrace
\end{align*}
Since the range of Heaviside function is only two values, namely $\{0,1\}$, we get the simple form for $g_2(\boldsymbol{\nu})$:
\begin{align*}
	g_2(\boldsymbol{\nu})  &= \sum_{i=1}^{N} q(\nu_i) \\
	\text{where} \quad  q(\nu_i) &=
	\begin{cases}
		(u_1 - u_0) \nu_i & \quad \text{if } \nu_i > 0 \\
		0 & \quad \text{otherwise}
	\end{cases} \\
	&= \left( u_1 - u_0 \right) \max(\nu_i , 0).
\end{align*}
This infimal value is attained at $\vphi^\star = H(\boldsymbol{\nu})$. Now the dual problem reads
\begin{align}
\begin{split}
	\underset{\boldsymbol{\nu} \in \mathcal{R}_{\mA}}{\min} \quad & \Big\lbrace \tfrac{1}{2} \| \boldsymbol{\nu} - \mA^T \vy \|_{\left( \mA^T \mA \right)^\dagger}^2 + \\
	& \sum_i \left( u_1 - u_0 \right) \max(\nu_i , 0) + u_0 \boldsymbol{\nu}^T \mathbf{1} \Big\rbrace.
	\label{eq:dualProblemGrey}
\end{split}
\end{align}
We note that the last two terms in the dual objective can be compactly represented by
\begin{align*}
	p(\boldsymbol{\nu}) = \sum_i |u_0| \max(-\nu_i,0) + |u_1| \max(\nu_i,0),
\end{align*}
where $p(\cdot)$ is known as an asymmetric one-norm. The optimal point of the problem \eqref{eq:dualProblemGrey} is denoted by $\boldsymbol{\nu}^\star$ and the corresponding primal optimal is retrieved using
\[
	\vx^\star = u_0 \mathbf{1} + (u_1 - u_0) H(\vphi^\star) = u_0 \mathbf{1} + (u_1 - u_0) H(\boldsymbol{\nu}^\star).
\]
\end{proof}

\subsection{Proposition \ref{thm:proxForAsyOneNorm}}
\begin{proof}
The minimization problem for the proximal operator of an asymmetric one-norm function $p(\cdot)$ reads
\begin{equation}
\underset{\vx \in \mathbb{R}^N}{\min} \quad f(\vx) = \tfrac{1}{2} \| \vx - \vz \|^2 + \lambda p(\vx), \label{eq:minProbAON}
\end{equation}
where $\lambda > 0$ is a parameter. Since the function is convex, we get the following from the first-order optimality condition \cite{rockafellar1970convex}:
\begin{align}
	\mathbf{0} & \in \partial f(\vx^\star), \nonumber \\
	& \in \vx^\star - \vz + \lambda \partial p(\vx^\star), \label{eq:optCondAON}
\end{align}
where $\vx^\star$ is an optimal point of \eqref{eq:minProbAON}, and $\partial p(\vx)$ is a sub-differential of function $p(\cdot)$ at $\vx$. This sub-differential is
\[
	\partial p(x_i) = \begin{cases}
	|u_1| & \quad x_i > 0 \\
	\left[ -|u_0|, |u_1| \right] & \quad x_i = 0 \\
	-|u_0| & \quad x_i < 0
	\end{cases}.
\]
Now coming back to the first-order optimality condition in \eqref{eq:optCondAON}, we get the explicit form for optimal solution $\vx^\star$:
\begin{align*}
	x_i^\star = \begin{cases}
		z_i - \lambda |u_1| & \quad z_i \geq \lambda |u_1| \\
		0 	& \quad -\lambda |u_0| \leq z_i \leq \lambda |u_1| \\
		z_i + \lambda |u_0| & \quad z_i \leq -\lambda |u_0|
	\end{cases}.
\end{align*}
We recognize this function as an asymmetric soft-thresholding function.
\end{proof}

\section*{Acknowledgment}
This work is part of the Industrial Partnership Programme (IPP) `Computational sciences for energy research' of the Foundation for Fundamental Research on Matter (FOM), which is part of the Netherlands Organisation for Scientific Research (NWO). This research programme is co-financed by Shell Global Solutions International B.V. The second author is financially supported by the Netherlands Organisation for Scientific Research (NWO) as part of research programme 613.009.032.


\begin{IEEEbiography}{Ajinkya Kadu}
received BSc. and MSc. degree in Aerospace Engineering from the Indian Institute of Technology, Bombay, India, in 2015. He is working towards the Ph.D. degree at Mathematical Institute of Utrecht University, The Netherlands. His Ph.D. is part of `Computational Sciences for Energy Research', a research programme of NWO in partnership with Shell. His research interests include computational imaging, full-waveform inversion and distributed optimization.
\end{IEEEbiography} 

\begin{IEEEbiography}{Tristan van Leeuwen}
received his BSc. and MSc. in Computational Science from Utrecht University. He obtained his PhD. in geophysics at Delft University in 2010. After spending some time as a postdoctoral researcher at the University of British Columbia in Vancouver, Canada and the Centrum Wiskunde \& Informatica in Amsterdam, the Netherlands, he returned to Utrecht University in 2014 as an assistant professor at the mathematical institute. His research interests include: inverse problems, computational imaging, tomography and numerical optimization.
\end{IEEEbiography}

\end{document}